\documentclass[11pt,a4paper]{article}
\pdfoutput=1
\usepackage{jcappub}
\usepackage{amsmath,amstext}
\usepackage{mathtools}
 \usepackage{commath}
\usepackage{bm}
\usepackage{graphicx}
\usepackage{multirow}
\usepackage{footmisc}

\usepackage{epsfig}
\usepackage{url}
\usepackage{hyperref}

\usepackage[normalem]{ulem}
\usepackage{epsfig}

\usepackage{amssymb}
\usepackage{graphicx}
\usepackage{verbatim}
\usepackage[usenames,dvipsnames]{xcolor}

\usepackage{subcaption}
\usepackage{adjustbox}
\newcommand{\bk}{{\bf k}}

\newcommand{\hiclass}{{\tt hi\_class}}

\newcommand{\lcdm}{{\Lambda{\rm CDM}}}

\newcommand{\bq}{\begin{eqnarray}}
\newcommand{\eq}{\end{eqnarray}}

\newcommand{\lp}{\emph{left panel}}

\newcommand{\rp}{\emph{right panel}}

\newcommand{\Aisw}{\mathcal{A}^{{\rm ISW}}}

\newcommand{\Abf}{\mathcal{A}^{bf}}

\newcommand{\fix}[1]{\textbf{ #1}}

\definecolor{grey}{rgb}{0.4,0.4,0.4}
\definecolor{dullmagenta}{rgb}{0.4,0,0.4}
\definecolor{darkblue}{rgb}{0,0,0.4}
\definecolor{midblue}{rgb}{0,0,0.5}
\definecolor{midred}{rgb}{0.5,0,0}
\definecolor{orange}{rgb}{1,0.5,0}
\definecolor{lightbrown}{rgb}{0.75,0.5,0.25}
\definecolor{tan}{cmyk}{0.14,0.42,0.56,0}
\definecolor{djunglegreen}{cmyk}{0.99,0,0.52,0}
\definecolor{lightgreen}{rgb}{0,1,0}
\definecolor{olivegreen}{cmyk}{0.64,0,0.95,0.40}
\definecolor{midgreen}{rgb}{0.0,0.675,0.0}
\definecolor{darkgreen}{rgb}{0,0.5,0}

\title{Galileon Gravity in Light of ISW, CMB, BAO and $H_0$ data}

\author[a]{Janina Renk}
\author[b,c]{Miguel Zumalac\'arregui}
\author[d]{Francesco Montanari}
\author[e]{Alexandre Barreira}

\affiliation[a]{The Oskar Klein Centre for Cosmoparticle Physics, Stockholm University \\ AlbaNova University Center, SE-106 91 Stockholm, Sweden}
\affiliation[b]{Nordita, KTH Royal Institute of Technology and Stockholm University \\ Roslagstullsbacken 23, SE-106 91 Stockholm, Sweden}
\affiliation[c]{Berkeley Center for Cosmological Physics, LBL and University of California at Berkeley, CA94720, USA}
\affiliation[d]{Physics Department, University of Helsinki and Helsinki Institute of Physics, \\
P.O. Box 64, 00014, University of Helsinki, Finland}
\affiliation[e]{Max-Planck-Institut f{\"u}r Astrophysik, Karl-Schwarzschild-Str. 1, 85741 Garching, Germany}

\emailAdd{janina.renk@fysik.su.se}
\emailAdd{miguelzuma@berkeley.edu}
\emailAdd{francesco.montanari@helsinki.fi}
\emailAdd{barreira@mpa-garching.mpg.de}

\abstract{
Cosmological models with Galileon gravity are an alternative to the standard $\lcdm$ paradigm with testable predictions at the level of its self-accelerating solutions for the expansion history, as well as large-scale structure formation. Here, we place constraints on the full parameter space of these models using data from the cosmic microwave background (CMB) (including lensing), baryonic acoustic oscillations (BAO) and the Integrated Sachs-Wolfe (ISW) effect.
We pay special attention to the ISW effect for which we use the cross-spectra, $C_\ell^{\rm T g}$, of CMB temperature maps and foreground galaxies from the WISE survey. The sign of $C_\ell^{\rm T g}$ is set by the time evolution of the lensing potential in the redshift range of the galaxy sample: it is positive if the potential decays (like in $\lcdm$), negative if it deepens.
We constrain three subsets of Galileon gravity separately known as the Cubic, Quartic and Quintic Galileons. The cubic Galileon model predicts a negative $C_\ell^{\rm T g}$ and exhibits a $7.8\sigma$ tension with the data, which effectively rules it out. For the quartic and quintic models the ISW data also rule out a significant portion of the parameter space but permit regions where the goodness-of-fit is comparable to $\lcdm$. The data prefers a non zero sum of the neutrino masses ($\sum m_\nu\approx 0.5$eV) with $ \sim \! 5\sigma$ significance in these models. The best-fitting models have values of $H_0$ consistent with local determinations, thereby avoiding the tension that exists in $\lcdm$. We also identify and discuss a $\sim \! 2\sigma$ tension that Galileon gravity exhibits with recent BAO measurements. Our analysis shows overall that Galileon cosmologies cannot be ruled out by current data but future lensing, BAO and ISW data hold strong potential to do so.
}


\date{\today}

\begin{document}

\maketitle

\section{Introduction} \label{sect:intro}

Theories of gravity beyond General Relativity (GR), commonly referred to as {\it modified gravity models}, have become the focus of growing attention in cosmological studies. The reason for this is mostly twofold. First, modifications to the gravitational law on large scales appear as a plausible alternative to the cosmological constant, $\Lambda$, to explain the accelerating expansion of the Universe. Second, analysing the predictions of modified gravity scenarios helps to understand the various types of observational signatures and therefore to improve the design of more robust tests of gravity on cosmological scales. Active research on modified gravity models has provided significant theoretical and observational advances over recent years (see e.g.~\cite{Jain:2007yk, 2012PhR...513....1C, Joyce:2014kja, 2015arXiv150404623K, 2016arXiv160106133J} for reviews).

The Covariant Galileon model \cite{PhysRevD.79.064036, Deffayet:2009mn, PhysRevD.79.084003} is a particularly interesting example of a theory of modified gravity. In this model, at the level of the background cosmology, the acceleration of the expansion of the Universe is driven by kinetic interactions of a scalar field whose Lagrangian density is invariant under the so-called Galilean shift symmetry $\partial_{\mu}\varphi \rightarrow \partial_{\mu}\varphi + b_{\mu}$, where $b_{\mu}$ is a constant four-vector and $\varphi$ is called the Galileon field (cf.~Eq.~(\ref{eq:Li}) below). At the level of perturbations the non-linear nature of the said kinetic interactions effectively couples the derivatives of the Galileon and metric fields together (in a process commonly dubbed as ``kinetic gravity braiding'' \cite{2011JHEP...11..156P, 2010JCAP...10..026D}). This changes the way gravitational potentials respond to density fluctuations which is why this model falls under the category of a modified theory of gravity that aims to explain ``dark energy'' (see e.g. \cite{Chow:2009fm,Gannouji:2010au,2010PhRvL.105k1301D,Appleby:2011aa} for the first few studies of cosmologies with Galileon gravity). One nice property of the Lagrangian of the Galileon model is that the same non-linearities that drive the acceleration of the expansion of the Universe are also responsible for ensuring that the model can pass the stringent Solar System tests of gravity that have been consistent with GR to very good precision \cite{Will:2014xja}. This occurs via a mechanism known as Vainshtein screening \cite{Vainshtein1972393, Babichev:2013usa, Koyama:2013paa} which effectively suppresses the size of the spatial gradient of the Galileon field (known as the \emph{fifth force}) in regions of high local matter density (see e.g.~\cite{Barreira:2013eea, Barreira:2013xea, Li:2013tda} and references therein, for studies of Vainshtein screening in the Galileon model). In this paper, we focus on length scales where linear theory holds and hence, the effects of screening do not come into play.

The Covariant Galileon model does not have a $\lcdm$ limit, i.e., there is no choice of model parameters for which the Galileon terms behave as a cosmological constant. This feature of the model implies that when confronted against observational data one is almost guaranteed to obtain best-fitting values of the cosmological parameters (e.g., the Hubble rate today $H_0$, cold dark matter density $\Omega_{cdm}h^2$, etc.) that are different from those in $\lcdm$. This means in particular that robust constraints on the model must come from analyses in which many (if possible all) cosmological parameters are allowed to vary to explore all possible degeneracies that may be at play. Such an extensive constraint analysis was carried out by \cite{Barreira:2014jha} in which the Galileon model was confronted against data from the full CMB temperature and lensing power spectrum as well as lower-redshift geometrical probes such as baryonic acoustic oscillations (BAO) data. In their results \cite{Barreira:2014jha} found that there are regions of the model parameter space that yield essentially the same goodness-of-fit to these data as $\lcdm$ (albeit with different best-fitting cosmological parameter values). Amongst the most interesting aspects of these constraints on Galileon gravity were (i) the best-fitting values of $H_0$, which (contrary to $\lcdm$) are automatically compatible with local Universe determinations; and (ii) the constraints on the summed neutrino masses $\Sigma m_\nu$ which are incompatible with $\Sigma m_\nu = 0$ with high significance (again, different than in $\lcdm$).

Another critical difference between Galileon cosmologies and $\lcdm$ concerns the evolution of the gravitational potentials. In \cite{Barreira:2014jha} the authors demonstrated that the modifications to gravity in the Galileon model are such that the lensing potential can deepen with time after matter domination, rather than strictly decaying as it is the case in standard $\lcdm$. This means that in $\lcdm$ the sign of the integrated Sachs-Wolfe (ISW) effect is always positive, whereas it can be negative in the Galileon model. The sign of the ISW effect in a specific redshift range can be probed by cross-correlating CMB temperature maps with the number counts of foreground galaxies (denoted $C_{\ell}^{\rm{Tg}}$ throughout) \cite{Giannantonio:2008zi}. Specifically, the amplitude and sign of this spectra is what is broadly referred to as the \emph{ISW amplitude} and \emph{ISW sign}. Various sources of recent evidence have been shown to be in agreement with $\lcdm$ in what concerns the sign of the ISW effect (see e.g.~\cite{Ade:2015dva}) which makes these data particularly useful to test the viability of the Galileon and other modified gravity models that strongly alter the behaviour of the gravitational potentials \cite{2009PhRvD..80f3536L, kimura2, 2014MNRAS.439.2978C, 2014MNRAS.442..821M, Enander:2015vja}. Data analyses from galaxy surveys correlated with the CMB temperature result in a $\sim 3 \sigma$ detection of a positive ISW amplitude (e.g.~\cite{Giannantonio:2008zi, Ferraro:2014msa,Ade:2015dva,Shajib:2016bes}). This implies that MG models that have a strictly growing lensing potential on sub-horizon scales can be ruled out with at least $3\sigma$ significance since they predict a negative ISW amplitude. The analysis of \cite{Barreira:2014jha} suggested that this could well be the case for their resulting best-fitting models but the discussion there was kept mostly qualitative. More quantitatively \cite{Renk:2016olm} confirmed that these best-fitting models indeed predict a negative $C_{\ell}^{\rm{Tg}}$ amplitude.

In this work, we aim to quantify more precisely the degree of the presumed tension of Galileon gravity with the ISW data. To do so, we carry out Monte Carlo Markov Chain (MCMC) explorations of the full parameter space in the Galileon model using CMB data from Planck (including lensing) \cite{Ade:2015xua} and BAO measurements%
\footnote{In the MCMC part of our analysis we consider the BAO compilation used in the Planck 2013 analysis: SDSS DR7 LRG \cite{padmanabhan20122}, BOSS DR9 CMASS \cite{anderson2012clustering} and the 6dF Galaxy Survey \cite{beutler20116df}. We do not include more recent BAO measurements as some of these are in tension with the best-fitting Galileon models to the CMB data which may prevent a consistent joint analysis \cite{Verde:2009tu}. Note however that we discuss these tensions separately in section \ref{sec:BAO-tens} by post-processing the MCMCs.}. %
We then calculate $C_\ell^{\rm{Tg}}$ for the accepted MCMC parameter space points to determine whether they are compatible with the ISW data. In our investigation we use the data obtained by cross-correlating CMB temperature maps with the galaxies from the \textit{Wide-field Infrared Survey Explorer} (\textit{WISE}) \cite{Wright:2010qw}; we shall refer to these data as the WISE ISW signal, for short. Our analysis will follow closely the steps described in \cite{Ferraro:2014msa, Shajib:2016bes} for $\lcdm$, but applied to the Galileon model and for a large range of parameter values. In particular, before evaluating $C_{\ell}^{\rm{Tg}}$, we first make use of the cross-correlation of CMB lensing maps with the galaxy distribution to get an estimate for the galaxy bias which would otherwise be completely degenerate with the Galileon effects on the amplitude of $C_{\ell}^{\rm{Tg}}$. One of our main results is that the ISW data do rule out a significant portion of the Galileon parameter space but leave behind regions that fit these data as well as $\lcdm$. Although we find that the Galileon model can pass current ISW tests we identify and discuss tensions with BAO measurements that were published after the constrain analysis of \cite{Barreira:2014jha}%
\footnote{From SDSS DR7 MGS \cite{Ross:2014qpa}, BOSS DR11 Ly$\alpha$-auto \cite{2015A&A...574A..59D}, BOSS DR11 Ly$\alpha$-cross \cite{Font-Ribera:2013wce} and BOSS DR12 Galaxy (combined LOWZ \& CMASS) \cite{Alam:2016hwk}.}. %
Our conclusion will be that the degree of tension with BAO is currently not strong enough to rule out the Galileon model but future higher precision data should confidently do so if the trend of current data gets confirmed.

The rest of the paper is organized as follows. In \autoref{sec:spectra} we list the equations used to evaluate the spectra that enter our analysis. \autoref{sec:gal} summarizes the main aspects of Galileon gravity and the current knowledge of its overall observational viability. Our main methodology steps are explained in \autoref{sec:met} and our results are shown in \autoref{sec:results}. We conclude with a summary and discussion in \autoref{sec:conc}.

\section{Theoretical Spectra and Observables} \label{sec:spectra}

In this section we present the equations used to predict theoretical power spectra for any metric theory of gravity in the linear regime of cosmological perturbation theory. We focus on the spectra relevant to the ISW part of the analysis. In all our results these spectra are evaluated using the {\hiclass} code\footnote{www.hiclass-code.net} \cite{2016arXiv160506102Z}. This code is a modified version of the CLASS code\footnote{www.class-code.net} \cite{2011arXiv1104.2932L} that follows the cosmology of general Horndeski theories of gravity \cite{Horndeski:1974wa}, of which the covariant Galileon is a particular example. We shall be brief in this section but refer the interested reader to \cite{Lesgourgues:2013bra,DiDio:2013bqa,DiDio:2016ykq} for a more complete account of the general expressions written in a form consistent with the conventions of the {\hiclass} code \cite{2016arXiv160506102Z}.

We consider scalar perturbations around a spatially flat Friedmann-Lema\^itre-Robertson-Walker (FLRW) spacetime in the longitudinal Newtonian gauge (we use natural units in which $c=1$):
\begin{equation} \label{eq:metric}
ds^2 = a^2 \left[ - \left( 1 + 2 \Psi \right) d\tau^2 + \left( 1 - 2 \Phi\right) \gamma_{ij} dx^i dx^j \right] \;,
\end{equation}
where $\tau$ is the conformal time, $a$ is the scale factor and $\Psi$ and $\Phi$ are the Bardeen potentials. The spatial part of the metric can be written as $\gamma_{ij} dx^i dx^j = \left[ dr^2 + r^2 \left( d\theta^2 + \sin^2 \theta d\varphi^2\right) \right] $.

The definition of the power spectrum of primordial curvature perturbations $\mathcal{R}(\bk)$ is given by $\langle \mathcal{R}(\bk)\mathcal{R}^{\star}(\bk') \rangle = P_{\mathcal{R}}(k) \delta_D(\bk-\bk')$ ($\delta_D$ is the Dirac delta function and the star denotes complex conjugation). It can be expressed in terms of the primordial amplitude $A_s$, pivot scale $k_{\rm pivot}$ and spectral index $n_s$. Given the coefficients $a_{\ell m}(z)$ of a spherical harmonics expansion at redshift z, the angular power spectra are defined as $\langle a_{\ell m}(z_i) a^{\star}_{\ell m}(z_j) \rangle = \delta_{\ell\ell'}\delta_{mm'}C_{\ell}(z_i,z_j)$, where $\delta_{\ell\ell'}$ is the Kronecker delta. They are given in terms of transfer functions $\Delta_{\ell}^{W_i}(k)$ as
\begin{equation} \label{eq:Cls}
C_{\ell}(z_i,z_j) = 4\pi \int \frac{dk}{k} \Delta_{\ell}^{W_i}(k) \Delta_{\ell}^{W_j}(k) \mathcal{P}_{\mathcal{R}}(k) \;.
\end{equation}

For calculations involving source number counts the relevant transfer function can be written as
\begin{equation}\label{eq:deltagterms}
\Delta_{\ell}^{{\rm g}_i} \approx \Delta_{\ell}^{\mathrm{Den}_i} +
\text{ other contributions},\;
\end{equation}
with
\begin{eqnarray}\label{eq:deltadens}
\Delta_{\ell}^{\mathrm{Den}_i} &=& \int_0^{\tau_0} d\tau W_i \, b_{\rm g}(\tau) \delta(\tau, k) j_{\ell}\;,
\end{eqnarray}
where we use $\delta(\tau, k)$ to denote the density perturbation at the Fourier mode $k$ and $j_{\ell}=j_{\ell}(x)$ with $x=k(\tau_0-\tau)$ are Bessel functions. Consistently with {\hiclass}, we consider all transfer functions to be normalized to the value of the curvature perturbation at some time $k\tau_{\rm ini} \ll 1$, e.g., $\delta(\tau,k)\equiv \delta(\tau,\bk)/\mathcal{R}(\tau_{\rm ini},\bk)$.

The terms not explicitly shown in Eq.~(\ref{eq:deltagterms}) (``other contributions'') encompass corrections from redshift-space distortions (RSD), lensing terms and contributions suppressed by $H/k$ that are small on sub-horizon scales (for the full computation of these terms, see \cite{Bonvin:2011bg,Challinor:2011bk,Yoo:2009au}). We have explicitly numerically verified that our results using the redshift distribution of the WISE sample are almost insensitive to the inclusion of the RSD and lensing terms\footnote{This conclusion holds for a range of typical values of the magnification bias parameters $s(z)$ \cite{Scranton:2005ci, 2017arXiv170207829L}.} (less than $1\%$ on all relevant scales and quantities). Hence, for all relevant practical purposes in this paper it is sufficient to include only the density term in Eq.~(\ref{eq:deltagterms}).

The bias of the WISE galaxies, $b_{\rm g}(z)$, enters the calculation via Eq.~(\ref{eq:deltadens}) and we assume it to be scale independent, since we are only considering scales $\ell < 400$. The selection function $W_i$ is given by the observed number of sources per solid angle and per redshift $\frac{dN}{dz d\Omega}$. We use the analytical approximation from \cite{Enander:2015vja} to the WISE galaxy selection function determined by \cite{Yan+}. This approximation (in arbitrary normalization) is given by 
\begin{equation} \label{eq:sel_fct}
\frac{dN}{dz d\Omega} = 61.3 - \frac{9.96}{0.142 + z} - 85 z \;,
\end{equation}
and set to zero where $\frac{dN}{dz d\Omega}(z) < 0$. This distribution is shown in the \lp\, of
\autoref{fig:Pot-ISW}.

In our analysis we follow the steps from e.g. \cite{Ferraro:2014msa,Ade:2015dva,Shajib:2016bes}, and use the CMB lensing convergence-galaxies cross-correlation $C_{\ell}^{\kappa {\rm g}}$ to fix the bias of the WISE galaxies. Then we calculate the CMB temperature-galaxy cross spectrum $C_{\ell}^{\rm Tg}$ to compare the Galileon predictions to the ISW data and assess the sign of the ISW effect\footnote{As commonly done in the literature (e.g. \cite{Ade:2015dva,Ferraro:2014msa,Shajib:2016bes,Enander:2015vja}) we use the term ``ISW Amplitude'' to refer to the sum of the spectrum $C_{\ell}^{\rm Tg}$ evaluated at a given set of multipoles $\ell$ (cf.~Eq.~(\ref{eq:ISWamp})).}. To compute these two spectra we need the transfer functions associated with the lensing convergence $\kappa$ and the ISW effects:
\begin{eqnarray}\label{eq:spectraused}
\Delta_{\ell}^{\kappa} &=& -\frac{\ell(\ell+1)}{2} \int_{\tau_*}^{\tau_0} d\tau \,
\frac{\tau-\tau_*}{(\tau_0-\tau)(\tau_0-\tau_*)} \,
\left(\Phi+\Psi\right) \,
j_{\ell} \;,\\
\Delta_{\ell}^{{\rm ISW}} &=& \int_{\tau_*}^{\tau_0} d\tau
\left(\Phi'+\Psi'\right) \, j_{\ell} ~.
\label{eq:delta_terms}
\end{eqnarray}
Here $\tau_*$ and $\tau_0$ are the conformal time at recombination and today, respectively, and we omitted the arguments $(k,\tau)$ for the transfer functions. Primes denote derivatives with respect to conformal time. The perturbation equations determining these transfer functions are solved numerically by {\hiclass} and are directly affected by the modifications of gravity \cite{2016arXiv160506102Z}. We stress that in {\hiclass} the transfer function of the lensing convergence is evaluated by using the definition of the convergence as the two-dimensional Laplacian of the lensing potential $\kappa=-\frac{1}{2}\Delta_{\Omega}\psi$. Given the direction of photon propagation ${\bf n}$, the lensing potential is given by \citep{Lewis:2006fu}
\begin{equation} \label{eq:lens_pot}
\psi({\bf n}, z) = - \int_0^{r_*}d{\tilde r} \frac{r_*-{\tilde r}}{r_*{\tilde r}} \left(\Phi+\Psi\right)({\tilde r}{\bf n}, \tau_0-{\tilde r}) \;.
\end{equation}
We do not use the popular relation $\kappa \propto \int {\rm d}\tau \delta$ since it is only valid in GR. Note also the dependence of $\Delta_{\ell}^{{\rm ISW}}$ on the time-derivatives of the Bardeen potentials, which is what characterizes the ISW effect. Correlations of CMB lensing convergence and temperature with low-redshift sources are well approximated, respectively, by
\begin{eqnarray}
C_{\ell}^{\kappa {\rm g}} &=& 4\pi \int \frac{dk}{k} \Delta_{\ell}^{\kappa}(k)
\Delta_{\ell}^{{\rm g}}(k) \mathcal{P}_{\mathcal{R}}(k) \;, \label{eq:kappag}
\\
C_{\ell}^{\rm Tg} &=& 4\pi \int \frac{dk}{k} \Delta_{\ell}^{{\rm ISW}}(k)
\Delta_{\ell}^{{\rm g}}(k) \mathcal{P}_{\mathcal{R}}(k) \;, \label{eq:Tg}
\end{eqnarray}
where we dropped the redshift index for number counts as it is now assumed that the galaxy transfer function is integrated over the whole respective redshift range of the galaxy sample. Finally, we note also that in the calculation of the spectra we make use of the Limber approximation \cite{Limber1953}. We numerically verified that the largest effect of applying this approximation is $\approx 1\%$ in the $C_{\ell}^{{\rm Tg}}$ and $C_{\ell}^{\kappa {\rm g}}$ spectra and hence, not a source of concern for our analysis here.


\section{Covariant Galileons}\label{sec:gal}
In this section we introduce the Galileon model and summarise the main aspects of its current observational status. We will be brief in our descriptions and refer the reader to the literature cited throughout for more details.

\subsection{Action of the Model}\label{sec:action}
The action of the minimally coupled covariant Galileon model is given by
\begin{equation}\label{eq:Li}
S[g_{\mu\nu},\phi]=\int\mathrm{d}^{4}x\,\sqrt{-g}\left[\sum_{i=2}^{5}{\cal L}_{i}[g_{\mu\nu},\phi]\,+\mathcal{L}_{\text{m}}[g_{\mu\nu},\psi_{\text m}]\right]\,,
\end{equation}
with
\begin{eqnarray}
{\cal L}_{2} & = & c_{2}X -\frac{c_{1}M^{3}}{2}\phi \,,\label{eq:L2}\\
{\cal L}_{3} & = & 2\frac{c_{3}}{M^{3}}X\Box\phi\,,\label{eq:L3}\\
{\cal L}_{4} & = & \left(\frac{M_{\rm Pl}^{2}}{2}+\frac{c_{4}}{M^{6}}X^{2}\right)R
+2\frac{c_{4}}{M^{6}}X\left[\left(\Box\phi\right)^{2}-\phi_{;\mu\nu}\phi^{;\mu\nu}\right]\,,\label{eq:L4}\\
{\cal L}_{5} & = & \frac{c_{5}}{M^{9}}X^{2}G_{\mu\nu}\phi^{;\mu\nu}
-\frac{1}{3}\frac{c_{5}}{M^{9}}X\left[\left(\Box\phi\right)^{3}+2{\phi_{;\mu}}^{\nu}{\phi_{;\nu}}^{\alpha}{\phi_{;\alpha}}^{\mu}-3\phi_{;\mu\nu}\phi^{;\mu\nu}\Box\phi\right]\,.\label{eq:L5}
\end{eqnarray}
In the above equations $g$ is the determinant of the metric $g_{\mu\nu}$, $R$ is the Ricci scalar, $G_{\mu\nu}$ is the Einstein tensor, $X\equiv-\partial_{\mu}\phi\partial^{\mu}\phi/2$, $\phi_{;\mu\nu} = \nabla_\mu \nabla_\nu\phi$, $\Box\phi = \nabla_\mu\nabla^\mu\phi$ and $\mathcal{L}_{\text m}$ denotes the Lagrangian of some matter field $\psi_{\text m}$. The mass scale $M^3\equiv M_{\rm Pl}H_0^2$ ensures that the $c_i$ coefficients remain dimensionless (where $M_{\rm Pl}$ is the Planck mass).

A few noteworthy points about the structure of the Lagrangian densities in Eq.~(\ref{eq:Li}) include:
\begin{itemize}
 \item The terms involving only first derivatives of the metric ($\mathcal{L}_2$, $\mathcal{L}_3$ and the terms in $\mathcal{L}_4$, $\mathcal{L}_5$ that do not involve $R$ or $G_{\mu\nu}$) represent all the possible terms whose equations of motion are invariant under a Galilean shift $\partial_\mu\varphi \rightarrow \partial_\mu\varphi + b_\mu$ and are kept up to second-order in field derivatives in four-dimensional Minkowski space \cite{PhysRevD.79.064036}.
 \item The explicit couplings to $R$ and $G_{\mu\nu}$ in $\mathcal{L}_4$ and $\mathcal{L}_5$ are not Galilean invariant but were included by \cite{Deffayet:2009mn} to keep the equations of motion second-order in fields derivatives in a spacetime like FLRW and hence, leave the theory free from instabilities known as Ostrogradski ghosts. Note however, that the addition of these couplings to curvature (which effectively act as counter terms that cancel higher-derivatives arising from the straightforward promotion of partial to covariant derivatives) is not necessary to yield ghost-free scenarios \cite{2014PhRvD..89f4046Z, 2015PhRvL.114u1101G,Langlois:2015cwa}.
 \item The model is minimally coupled to matter, i.e. the matter Lagrangian is constructed out of $g_{\mu\nu}$ and matter $\psi_{\text m}$ fields with no explicit occurrence of $\phi$. This is different from other models which often feature a coupling to the matter energy-momentum tensor (see e.g.~\cite{Bettoni:2013diz,Ezquiaga:2017ner} for the explicit Lagrangian).
\end{itemize}

Below, we focus specifically on the model of Eq.~(\ref{eq:Li}). However, analyses such as the one we perform here should also result in powerful constraints on other versions of the Galileon model, including the beyond Horndeski covariantization \cite{2015PhRvL.114u1101G,Gleyzes:2014qga} and models with non-minimal couplings to matter, in the sense that these models also generally exhibit non-trivial evolutions of the lensing potential. This behaviour can be traced to generic properties of modified gravity theories \cite{Renk:2016olm} that are present in all variants of Galileon gravity \cite{Bellini:2014fua,Gleyzes:2014qga}.

\subsection{The Galileon Subspace of Parameters} \label{sec:gal-bg}

The dimensionless constants $c_{1-5}$ are model parameters to be constrained by observational data. The parameter $c_1$ describes a linear potential term with no particularly interesting dynamics and hence we will set it to zero. A rescaling of the scalar field $\phi$ by a constant factor $B$, $\phi \rightarrow B\phi$, preserves the physics of the model as long as the Galileon parameters are rescaled as $c_i \rightarrow c_i/B^i$. If this unphysical degeneracy is not broken, then the Markov chains cannot converge (see \cite{2013PhRvD..87j3511B} for a more detailed discussion). To break the degeneracy, we follow \cite{Barreira:2014jha} and fix $c_2 = -1$. This way, the $\mathcal{L}_2$ piece reduces to the standard scalar kinetic term with a negative sign. The fact that $c_2<0$ is an observational requirement \cite{2013PhRvD..87j3511B, Barreira:2014jha} and does not necessarily lead to ghost-like instabilities. We stress that the physics of the model are not affected by which parameter we choose to fix (for instance, \cite{2013PhRvD..87j3511B} fixed $c_3=10$ and quoted constraints on combinations such as $c_4/c_3^{4/3}$), or in other words, the constraints and best-fitting values we obtain here can be straightforwardly translated to other choices of the fixed parameter analytically.

When solving for the background evolution one must specify the initial condition of the Galileon field time derivative $\dot{\phi} = {\rm d}\phi/{\rm d}t$ ($t$ is the physical time), which is in general a free parameter. In \cite{2013PhRvD..87j3511B} the authors found that in order for the model to yield satisfactory fits to the CMB data, the background in the Galileon model must reach the so-called tracker evolution before the energy density of the Galileon field starts to contribute non-negligibly to the total energy density of the Universe (see e.g.~Fig.~11 of \cite{Barreira:2014jha}). The results are insensitive to the exact time this tracker evolution starts to be followed, provided that the tracker is reached before this critical time. This evolution is characterized by \cite{2010PhRvL.105k1301D}
\begin{equation}\label{eq:xi_def}
\frac{\dot{\phi}H}{M_{\text Pl} H_{0}^{2}} \equiv \xi = {\rm constant}\, ,
\end{equation}
%
where $\xi$ is a constant free dimensionless parameter. Below, we assume that the Galileon field is always on the tracker, which effectively means setting up the initial condition of $\dot{\phi} = {\rm d}\phi/{\rm d}t$ to satisfy Eq.~(\ref{eq:xi_def}). 
If Galileon dynamics are valid during inflation, then it is interesting to note that they are naturally set extremely close to the tracker value in the early universe. The existence of the tracker is a direct consequence of shift symmetry ($\phi\to\phi+c$), by virtue of which the evolution of the field is equivalent to the covariant conservation of the shift-current \cite{Bellini:2014fua} $\nabla_\mu{\cal J}^\mu\Rightarrow \dot{\cal J}^{0} + 3H{\cal J}^0=0$, where ${\cal J}^0$ is defined in Eq.~(\ref{eq:constraints}). The general solution ${\cal J}^0 \propto a^{-3}$ decays towards zero as the inverse volume of the universe. It can hence be expected to have dropped to a negligible value by the end of inflation $\frac{a_{\rm end}}{a_{\rm ini}}\sim e^{50-60}$, or deep in the radiation era when \texttt{hi\_class} initial conditions are set.%
\footnote{In general, without assuming the tracker evolution, one could have allowed $\dot{\phi}_i = {\rm d}\phi_i/{\rm d}t$ to be a free parameter but the data will only put an upper bound on it (see \cite{2013PhRvD..87j3511B} for an analysis varying the initial condition). Below this upper bound the predictions are then always the same, which is why we can assume the tracker evolution at all times. Recently in \cite{2017A&A...600A..40N}, the authors have placed constraints on the Galileon model using cosmological data and found that the tracker evolution is less favoured by the data compared to more general background evolutions. In \cite{2017A&A...600A..40N} however, the constraints do not include data from the amplitude of the CMB temperature power spectrum which plays the dominant role in setting constraints on the Galileon model. The authors in \cite{2017A&A...600A..40N} also did not consider the impact of massive neutrinos which play an important role in the background expansion in the Galileon model (see e.g.~\cite{Barreira:2014jha}).}

The functional form of the Hubble rate in the tracker solution of the Galileon model is given by
\begin{equation}\label{eq:hubletracker}
\left(\frac{H(a)}{H_0}\right)^2 = \frac{1}{2}\Bigg[\Omega_{\rm fluid}(a) + \sqrt{[\Omega_{\rm fluid}(a)]^2 + 4\Omega_{\phi 0}}\Bigg],
\end{equation}
with which one can notice the lack of a $\lcdm$ limit. Here, $\Omega_{\rm fluid}$ is the energy density of matter and radiation and $\Omega_{\phi 0}$ the fractional energy density of the Galileon field today.
The condition that the evolution is that of the tracker (\ref{eq:xi_def}), together with the condition for the Universe to be spatially flat (derived from Eq.~(\ref{eq:hubletracker})), result in the two following constraints \cite{Barreira:2014jha}:
\bq\label{eq:constraints}
\Omega_{\phi 0}&=&\frac{c_{2}}{6}\xi^{2}-2c_{3}\xi^{3}+c_{4}\frac{15}{2}\xi^{4}+c_{5}\frac{7}{3}\xi^{5}, \nonumber \\
\mathcal{J}^0 &=& c_{2}\xi-6c_{3}\xi^{2}+18c_{4}\xi^{3}+5c_{5}\xi^{4} = 0.
\eq
These constraints allow to fix two Galileon parameters in terms of the others, effectively reducing the dimensionality of the Galileon subspace of parameters by 2 (recall that $c_2 = -1$).

To organize our discussions below we divide the Galileon subspace of parameters into three sectors of increasing complexity:
\begin{enumerate}
 \item Cubic Galileon ($Gal 3$): $c_4=c_5=0$ such that $c_3$ and $\xi$ are fixed by Eqs.~(\ref{eq:constraints}). This model has the same number of free parameters as standard $\lcdm$.
 \item Quartic Galileon ($Gal 4$): $c_5 = 0$, $c_3$ and $c_4$ fixed by Eqs.~(\ref{eq:constraints}), with $\xi$ left as a free parameter. This model has one extra parameter relative to $\lcdm$.
 \item Quintic Galileon ($Gal 5$): $c_4$, $c_5$ fixed by Eqs.~(\ref{eq:constraints}), such that $c_3$ and $\xi$ are left as free parameters. This is the most general case we consider which has two extra parameters relative to $\lcdm$.
\end{enumerate}
In our analysis we always require all models to be free from ghost or Laplace instabilities on scalar and tensor perturbations. These stability criteria (see e.g.~\cite{2016arXiv160506102Z}) depend only on the background evolution and therefore can be checked before solving for the evolution of the perturbations in $\hiclass$.

\subsection{Observational Status of the Galileon Model and the ISW Effect}\label{sec:status}
To the best of our knowledge the latest thorough account on the observational status of the covariant Galileon model is that of \cite{Barreira:2014jha}. There, the authors placed observational constraints on the full (cosmological + Galileon) parameter space of the model using the CMB temperature and CMB lensing data products from Planck 2013 \cite{Ade:2013sjv} as well as a compilation of BAO scale measurements at lower redshift%
\footnote{In the analysis of \cite{Barreira:2014jha} the authors have also measured the impact of using SNIa data but such results are not shown given that they had a sub-dominant impact on the overall constraints and conclusions.} %
that includes SDSS DR7 LRG \cite{padmanabhan20122}, BOSS DR9 CMASS \cite{anderson2012clustering} and the 6dF Galaxy Survey \cite{beutler20116df}.
We refer the reader to Section IV of \cite{Barreira:2014jha} for a summary of other past observational constraint analyses performed on the Galileon model.%
\footnote{Galileon gravity can also be constrained by local and astrophysical tests of gravity. Spherically symmetric solutions linear in time $\phi_{\rm local}(t,r) = \dot\phi(t_0)\cdot t + \varphi(r) + \phi(t_0)$ lead to strong constraints on the quartic and quintic models, i.e. through time variation of the gravitational constant \cite{Babichev:2011iz} or the orbital decay of binary pulsars \cite{Jimenez:2015bwa} if the local and cosmological time derivative of the scalar field have similar values $\dot\phi_{\rm local}(t,\vec x) \approx \dot{\phi}_{\rm cosmo}(t)$. This property of the local solution has been derived for shift-symmetric theories assuming that $\ddot{\phi}(t) = 0$ on the cosmological solution \cite{Babichev:2011iz}. For Galileon gravity this statement is true asymptotically in the de Sitter limit cf. (\ref{eq:xi_def},\ref{eq:hubletracker}). However, the second derivative in the attractor solution today
\begin{equation}
 \ddot\phi_{\rm cosmo} = \frac{\xi}{H_0^{2}} \frac{3 \Omega_m}{2(1+\Omega_{gal})}\approx \frac{0.25}{\xi}\dot\phi^2_{\rm cosmo}\,,
\end{equation}
is non-negligible in general and introduces new terms that can affect the local solution. The quartic Galileon equation (A1 of Ref. \cite{Bettoni:2015wta}) includes terms like $G^{\mu\nu}\phi_{;\mu\nu}\sim \ddot\phi R^{00}$ that vanish if $\ddot\phi=0$: these terms are enhanced in the local solution, as $R_{00}^{\rm local} \propto \rho_{\rm local} \gtrsim 10^{30}\rho_{\rm cosmo}$ for the Earth and $10^{44}$ for a neutron star in a binary pulsar. For those reasons, the connection between cosmological and local solutions in these theories does require more detailed modeling than currently available. We proceed with focus on cosmology, but note that these are considerations that should be revisited given their potentially critical importance in setting the observational viability of the Galileon model.}

Overall, in \cite{Barreira:2014jha} it was found that there are regions in the cosmological parameter space of the Galileon model that yield the same goodness-of-fit as $\lcdm$ to the CMB temperature and lensing power spectrum, as well as to the BAO data. A caveat that was pointed out, however, is that the typical best-fitting models exhibit a growth of the lensing potential at late times, which is in contrast with the well-known decay predicted by $\lcdm$. In the latter model the decay of the lensing potential is caused by the onset of the accelerated expansion of the Universe. In the case of the best-fitting Galileon models the effects of the acceleration on the lensing potential are insufficient to counteract the fast growing impact of the Galileon field which can work very effectively to make the potentials deeper closer to the present-time on large scales (cf.~Figure 3 and 7 of \cite{Barreira:2014jha}). 

If the lensing potential decays with time, as it does in standard $\lcdm$, then the ISW effect causes a positive cross-correlation between the CMB temperature and the foreground distribution of matter. This positiveness of the ISW effect is in line with various pieces of observational evidence \cite{Ho:2008bz,Ade:2015dva}. If a negative ISW effect is a general prediction of the Galileon model, then the ISW signal holds a great potential to rule out this theory of gravity or at least place very tight constraints on it. In \cite{Barreira:2014jha} the authors demonstrated that the lensing potential grows with time for a few best-fitting Galileon models, and limited themselves to arguing qualitatively that the ISW effect should be negative and, as a result of that, these models would be ruled out. The work of \cite{Renk:2016olm} has subsequently confirmed that the said best-fitting models had indeed a negative cross-correlation for galaxies distributed around $z \approx 0.3$. The effects of the non-linear Vainshtein screening mechanism do not help at easing these observational tensions because the ISW signal is sensitive almost exclusively to linear structure formation processes on scales $\gtrsim 10\ {\rm Mpc}/h$, whereas the screening effects only become important on scales $\lesssim 1-10\ {\rm Mpc}/h$ (see e.g.~\cite{Barreira:2013eea, Li:2013tda, Barreirahalomodel}).

\begin{figure}
 \centering
 \begin{minipage}{0.49\linewidth}
    \centering
   {\includegraphics[width=\textwidth]
    {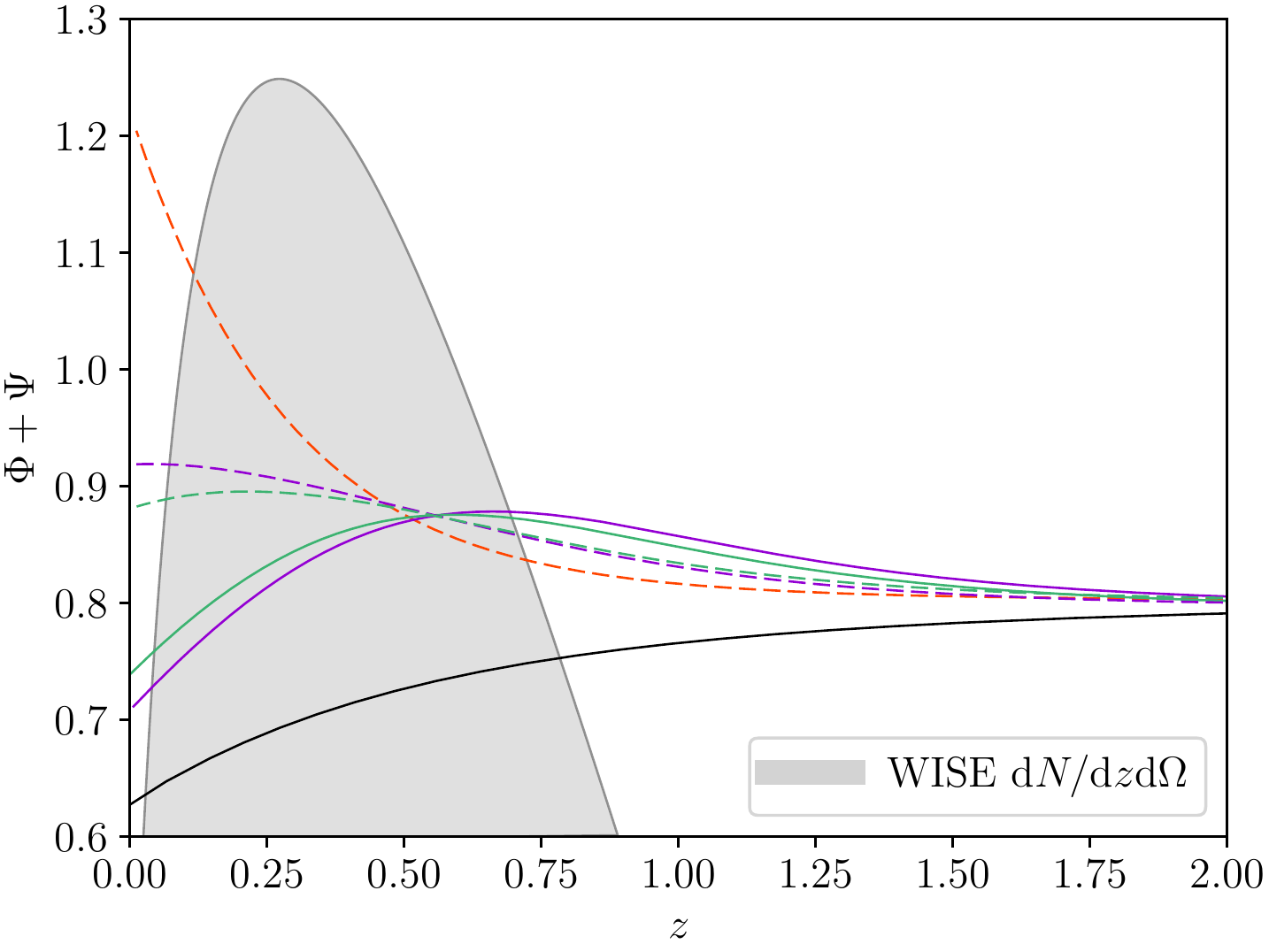}} \\
 \end{minipage}
 \hfill
 \begin{minipage}{0.49\linewidth}
    \centering
	{\includegraphics[width=\textwidth]
	{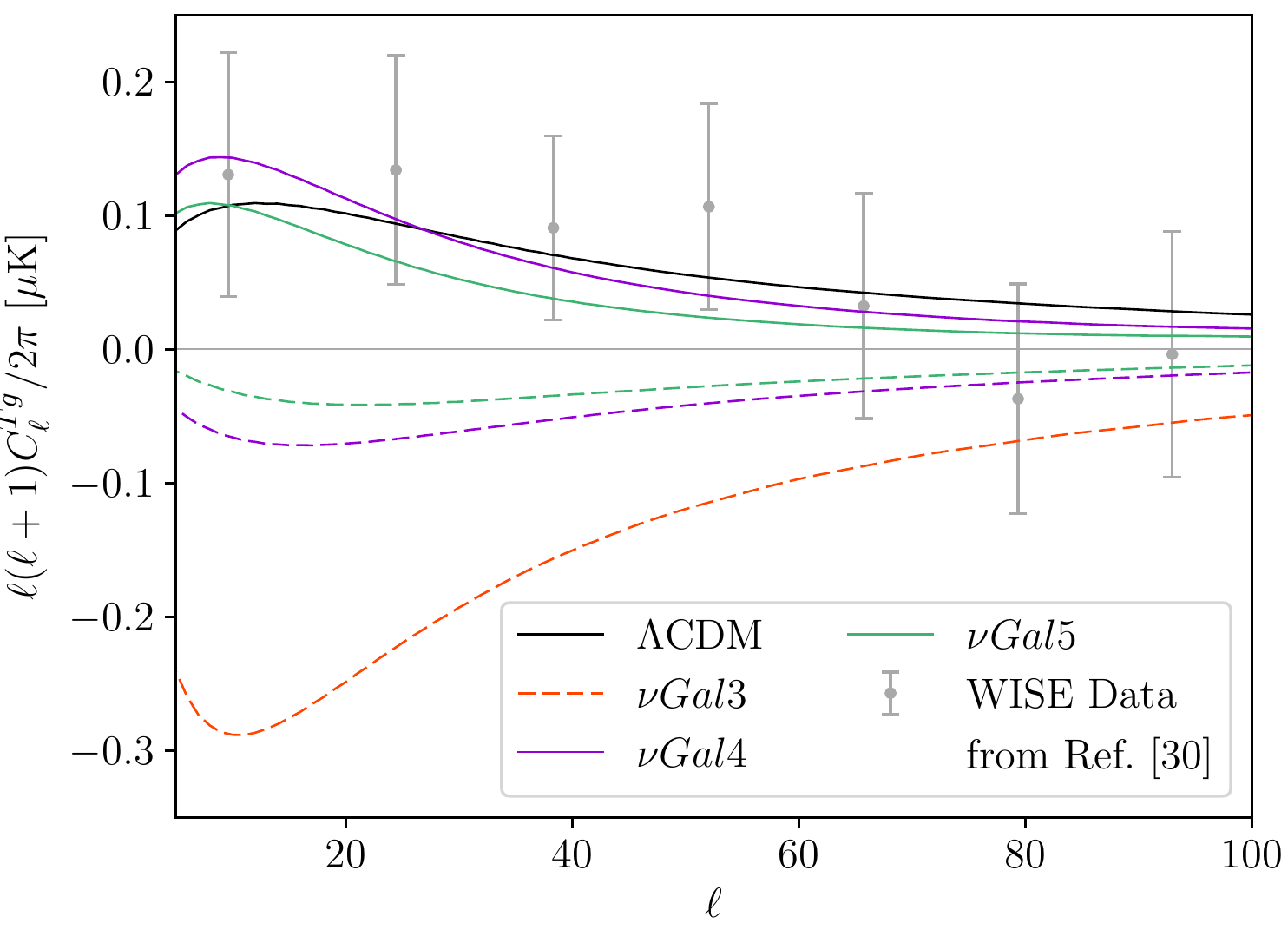}} \\
 \end{minipage}
\caption{Lensing potential on scale $k = 0.01 /$Mpc as a function of redshift (\lp) in code units of CLASS and CMB temperature - WISE galaxy cross-correlation (\rp) for $\Lambda$CDM and the Galileon models. The shaded region in the \lp \ indicates the WISE redshift selection function $\mathrm{d}N/\mathrm{d}z \mathrm{d} \Omega$ given in Eq.~(\ref{eq:sel_fct}) with adjusted offset and normalization for display. The black solid line shows the prediction of $\lcdm$ while the coloured dashed/solid lines indicate examples of Galileon models with growing/decaying potentials within the redshift range of the WISE selection function. Cubic models are shown in orange ($\nu Gal3$), quartic in purple ($\nu Gal4$) and quintic ($\nu Gal5$) in green. The temperature-galaxy data are the Q-band measurements from \cite{Ferraro:2014msa}.}
\label{fig:Pot-ISW}
\end{figure}

The Galileon subspace of parameters, however, provides enough freedom to obtain evolutions of the lensing potential that yield a positive ISW effect. This is illustrated in \autoref{fig:Pot-ISW}. The \lp \ shows the redshift evolution of the lensing potential for $\lcdm$ and representative Galileon models, as labelled. The shaded region depicts the redshift distribution of the WISE galaxies we use in this paper. For the Galileon curves shown the dashed ones correspond to cases with growing lensing potentials. The \rp \ shows the resulting $C_{\ell}^{\rm{Tg}}$ spectrum which is negative and, hence, at odds with the WISE ISW data (grey points). On the other hand there are choices of the Galileon parameters that yield decreasing potentials (solid lines). An interesting point to note for these curves is that, although the potential can grow in some redshift ranges (e.g.~$z\sim0.5-1$) it is decaying in the redshift range spanned by the WISE galaxies. This therefore yields a positive $C_{\ell}^{\rm{Tg}}$, as shown in the \rp. A main question that we address below is then: {\it is the positiveness of the ISW effect in Galileon cosmologies compatible with CMB and BAO data?} We will see below that yes: there are regions in the parameter space that yield an acceptable fit to the CMB, BAO and ISW data considered in this paper%
\footnote{There are some recent BAO scale determinations that are in tension wit Galileon gravity which we discuss in \autoref{sec:BAO-tens}.}.

%

%
%

\section{Methodology} \label{sec:met}
In this section we outline the main steps taken in our analysis. The first step consists in placing constraints on the parameter space of the Galileon model using data from the CMB and BAO. This serves to pin down the parts of the parameter space that merit the subsequent dedicated ISW analysis. The latter involves first calibrating the bias using data from the $C_{\ell}^{\kappa {\rm g}}$ spectrum and then analysing the predicted $C_{\ell}^{{\rm T g}}$ spectrum to assess the goodness-of-fit to the ISW data.

\subsection{MCMC constraints}\label{sec:mcmc}
The MCMC exploration of the Galileon model parameter space is carried out using the \hiclass \, and MontePython \cite{Audren:2012wb} codes. We use data from the CMB temperature power spectrum and CMB lensing potential power spectrum from Planck (temperature, polarization and lensing) \cite{Ade:2015xua} as well as BAO measurements from SDSS DR7 LRG \cite{padmanabhan20122}, BOSS DR9 CMASS \cite{anderson2012clustering} and the 6dFGS \cite{beutler20116df}%
\footnote{The likelihoods we have used in MontePython are '{\texttt Planck\_highl}', '{\texttt Planck\_lowl}', '{\texttt Planck\_lensing}' and '{\texttt bao\_boss}' with the data points as stated above.}. %
These BAO measurements are those which were used in the cosmological constraint analysis of Planck 2013 \cite{Ade:2013zuv} and to constrain the Galileon Models in \cite{Barreira:2014jha}. Admittedly, there are more recent BAO scale determinations but as we will discuss in \autoref{sec:BAO-tens} they are in some tension with the Galileon models. Hence, to avoid risking having biased best-fitting regions we opt to run the MCMC analysis with a BAO compilation that is not in tension with the Galileon model (but that provides with enough constraining power), and then subsequently check the goodness-of-fit of the best-fitting regions to more recent BAO data. For short, we will refer to this dataset as CMB+BAO13. 

We place separate constraints on the cubic, quartic and quintic Galileon models. For all these models we vary the following cosmological parameters (in addition to the relevant Galileon parameters; cf.~\autoref{sec:gal-bg}):
\bq\label{eq:params}
\Big\{100\omega_b, \omega_{cdm}, H_{0}, n_s, 10^{9}A_{s}, \tau_{reio}, \Sigma m_\nu\Big\},
\eq
which are, respectively, the physical baryon matter density $\omega_b = \Omega_{b0}h^2$, the physical cold dark matter density $\omega_{cdm} = \Omega_{cdm 0}h^2$, the Hubble rate today $H_0 = 100h\ {\rm km/s/Mpc}$, the scalar spectral index of the primordial power spectrum $n_s$, its amplitude $A_s$ at a pivot scale $k_{\rm pivot} = 0.05\ {\rm Mpc}^{-1}$, the optical depth to reionization $\tau_{reio}$, and the summed mass of three active neutrinos $\Sigma m_\nu$. Neutrino masses are in general an unknown parameter that should be varied and constrained by the data; while direct searches provide model-independent determinations of the total neutrino mass, the constraints are still not informative enough for our purposes ($0.06 {\rm eV} < \Sigma m_{\nu} < 6.6 {\rm eV}$ \cite{Drexlin:2013lha}). Rather than assuming a specific, possibly biased prior, we take a more general approach and only require neutrino masses to be non-negative. In the case of the Galileon model they even play a fundamental role in providing acceptable fits to the data \cite{Barreira:2014jha}. For this reason we explicitly include the symbol $\nu$ into the Galileon model abbreviations ($\nu Gal 3$, $\nu Gal 4$ and $\nu Gal 5$) to emphasize that neutrino masses are a free parameter (which is sometimes neglected in observational constraint analyses). For all models we always consider a degenerate mass spectrum for the three families of active neutrinos. Furthermore, despite the stability conditions on the Galileon parameters (see \autoref{sec:gal-bg}) we impose uninformative priors.

This part of our analysis consists essentially in an update and validation of the analysis done in \cite{Barreira:2014jha}. These chains are then sampled to check the compatibility with the ISW data as outlined next. We do not include the ISW data directly into the MCMC exploration due to computational costs: the precision needed to compute the CMB temperature-galaxy distribution cross-spectrum slows down the computation considerably. To obtain converged chains within a reasonable time limit we only include CMB+BAO data into the MCMC analysis. Having a converged set of chains, we then downsample them by a factor of $\sim \!10$ to inspect how  the ISW amplitude varies across the parameter space allowed by the CMB+BAO data.

\subsection{Galaxy Bias Calibration} \label{sec:bias}
The amplitude of the cross-correlation of the CMB temperature with galaxies, $C_\ell^{{\rm Tg }}$, is degenerate with the bias of the galaxies which enters through $\Delta^g_\ell(k)$ in Eq.~(\ref{eq:Tg}). Therefore one has to estimate the bias of the WISE galaxies first before assessing the impact of the ISW data on the observational viability of the Galileon model. Here, we follow similar steps as in \cite{Ferraro:2014msa, Ade:2015dva, Shajib:2016bes} and use the cross-correlation between the CMB lensing potential and the galaxy number counts, $C^{\kappa {\rm g}}_{\ell}$, on scales $100 \lesssim \ell \lesssim 400$ to fit for the bias.

In Eq.~(\ref{eq:kappag}), we use linear theory to compute the transfer functions and consider a simple redshift-dependent bias with $b(z) = b_0 (1+z)$, as done by the Planck Collaboration \cite{Ade:2015dva} to estimate the WISE galaxy bias. Naturally, this treatment can be made more robust (e.g. inclusion of non-linearities, and eventual scale-dependence of the bias), but this simple modelling is sufficient for our purposes here. For any given point in parameter space analysed we fit for the value of $b_0$ by maximizing the following Gaussian likelihood function
\begin{equation}
\mathcal{L} (\bm{d};\bm{t(b)},\mathrm{C}) \propto \exp \left[-\frac{1}{2}(\bm{d}-\bm{t})^T \mathrm{C}^{-1} (\bm{d}-\bm{t})\right],
\end{equation}
where $\bm{t}$ is the theoretical prediction which is evaluated according to Eq.~(\ref{eq:kappag}). The data vector, $\bm{d}$, and the associated covariance, $\mathrm{C}$, are both taken from \citep{Ferraro:2014msa}; in the latter the authors use the reconstructed lensing potential map from the Planck 2013 results \cite{Ade:2013sjv}%
\footnote{The update to the lensing potential map from Planck 2015 \cite{Adam:2015rua} would only lead to a minor decrease of the bias; see Fig. 7 of \cite{Ferraro:2016ymw}.}%
, the CMB temperature map of the WMAP 9-year results \cite{Bennett:2012zja}%
\footnote{Differences between the Planck 2015 and the WMAP 9-year CMB temperature map are negligible for the purpose of this analysis on scales $ \ell \lesssim 100$.}%
 and the galaxy catalogue from the WISE survey \cite{Wright:2010qw}. 

\begin{figure}
 \centering
 \begin{minipage}{0.6\linewidth}
    \centering
   {\includegraphics[width=\textwidth]
   {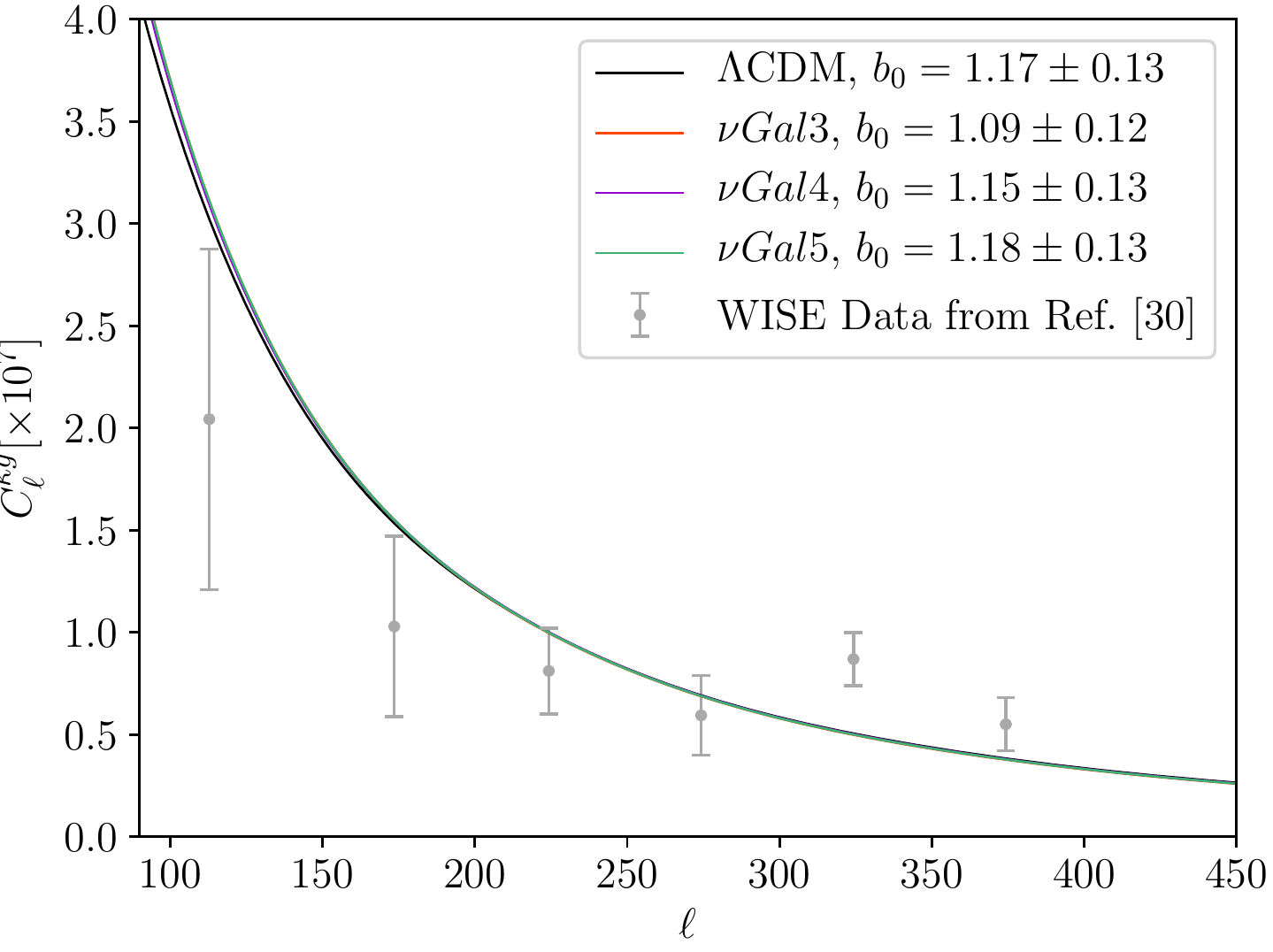}} \\
 \end{minipage}
 \caption{Lensing convergence-galaxy cross-correlation for $\lcdm$ and for a representative cubic, quartic and quintic model with the best-fitting values of $b_0$ for the redshift dependent bias model $b(z)= b_0 (1+z)$ indicated in the legend; note that the Galileon curves are overlapping and indistinguishable. The data points are from \cite{Ferraro:2014msa}.}
 \label{fig:kappa-g_zdep}
\end{figure}

\autoref{fig:kappa-g_zdep} shows the best-fitting $C_\ell^{\kappa \rm{g}}$ for $\lcdm$ and for representative cubic, quartic and quintic Galileon models. The agreement between the theoretical spectra and the data is not perfect. For the case of $\lcdm$, better fits are obtained if instead of using linear theory to evaluate the transfer function in Eq.~(\ref{eq:kappag}) one uses non-linear prescriptions like {\tt Halofit} \cite{Smith:2002dz, 2012ApJ...761..152T}. This reduces the best-fitting value of $b_0$ by about $25\%$, which translates into a decrease in the amplitude of the ISW signal ($\Aisw$ in Eq.~(\ref{eq:ISWamp}) below) of the same order. Such non-linear prescriptions are not available for Galileon gravity and thus we have to rely on linear theory. We note, however, that for our goals in this paper it is sufficient to determine only roughly the value of $b_0$ to break the degeneracy with the effects of the Galileon field on the amplitude of $C_\ell^{{\rm Tg}}$. Our main conclusion, that Galileon cosmologies admit good fits to ISW data, is not sensitive to the exact value of $b_0$, but the precise best-fitting Galileon parameter values are.

\subsection{Fit to ISW Data} \label{sec:ISW}
Finally, the last step in the search for Galileon models consistent with ISW data concerns the actual calculation of the cross-correlation between the CMB temperature and the WISE galaxies, $C_\ell^{{\rm Tg }}$. We define the amplitude of the ISW effect $\Aisw$ as 
\begin{equation}\label{eq:ISWamp}
 \Aisw = \frac{\sum_i {t}_i}{\sum_i {d}_i},
\end{equation}
where $\bm{t}$ and $\bm{d}$ are the vectors containing the theoretical predictions and the data as measured in \cite{Ferraro:2014msa}, respectively, and the index $i$ runs over the multipoles of the data (cf.~\autoref{fig:Pot-ISW}). The sign of $\Aisw$ provides a quick diagnostic of the overall sign of the ISW (provided the spectra does not oscillate non-trivially around zero), but it is not very informative about the overall goodness-of-fit to the data. To determine this we also compute the following $\chi^2_{\rm ISW}$ quantity
\begin{equation}
	\chi^2_{\rm ISW} = (\bm{d}-\bm{t})^T \mathrm{C}^{-1}
	(\bm{d}-\bm{t}) \;,
\end{equation}
where $\mathrm{C}$ is the covariance matrix from \cite{Ferraro:2014msa}.

To determine the level of agreement/tension of a given model we can calculate the best-fitting amplitude, $\Abf$, by rescaling the theoretical prediction as $\bm{t}\to f\bm{t}$ and then fit $f$ to the data. The minimization $\dif \chi^2 / \dif f = 0$ can be carried out analytically to yield
\begin{equation} \label{eq:Abf}
	\Abf = \bm{d}^T \mathrm{C}^{-1} \bm{t} / \left(\bm{t}^T \mathrm{C}^{-1} \bm{t}\right) \;,
\end{equation}
with variance
\begin{equation} \label{eq:sigmabf}
	\sigma_{bf}^2 = 1 / \left(\bm{t}^T \mathrm{C}^{-1} \bm{t}\right) \,.
\end{equation}
For a given model, the value of $|A^{bf}-A^{ISW}|/\sigma_{bf}$ gives a measure of the level of agreement between theory and observations.

The calculation of the WISE ISW signal requires knowledge of the redshift selection function of the survey. We use Eq.~(\ref{eq:sel_fct}) as an analytical approximation to the selection function originally put forward in \citep{Yan+} (see their Fig.~4). In the Galileon model the evolution of the lensing potential and consequently the ISW signal may depend sensitively on the redshift range analysed, as can be seen in the \lp\ of \autoref{fig:Pot-ISW}. We numerically verified that the use of the analytical approximation compared to the distribution determined by \cite{Yan+} does not affect our results by more than $1\%$ in the best-fitting $b_0$ or $\Aisw$ values. %
To account for the uncertainties in the precise redshift range spanned by the WISE galaxies we follow the steps from \cite{Ferraro:2014msa} who shifted the whole selection function by $\Delta z = \pm 0.1$ as a test. In \cite{Ferraro:2014msa}, the authors found that in $\lcdm$ the change of $\Abf$ is only about $5\%$ corresponding to $\sim 15\% \sigma_{bf}$; as expected from the evolution of the lensing potential shown in \autoref{fig:Pot-ISW} the effect on the Galileon amplitudes are more significant and can result in a change of $\Abf$ of $\sim 20\%$. Nevertheless, we verified that our general conclusions are unaffected by this: the change of the central redshift of the selection function in the Quintic Galileon model results only in an effective shift in the Galileon parameter space of the models that provide a good fit to the ISW data. Owing to this degeneracy the overall goodness-of-fit to the CMB+BAO13+ISW data remains unaffected compared to the unshifted distribution. 
This makes us confident that our overall conclusions are not dependent on the exact modelling of the redshift distribution of the WISE galaxies. Although we note that precise determinations of bounds on best-fitting Galileon parameters may be specific to the precise modelling of the redshift distribution function.


\section{Results} \label{sec:results}
In this section we present the results of the methodology outlined in the previous section for Cubic, Quartic and Quintic Galileons.

\begin{table*}
\centering
\begin{tabular}{@{}lccccccccccc}
\hline\hline
\\
 & \ \ $\lcdm$ \ \ & $\nu Gal 3$ & \ \ $\nu Gal 4$ \ \ & $\nu Gal 5$ & \\
\hline \hline
\\
$100\omega_b$  			 &\ \ $2.222^{+0.041}_{-0.039}   $ 	 & \ \ $2.199^{+0.041}_{-0.040}   $	&\ \ $2.205^{+0.040}_{-0.039}   $	&\ \ $2.204^{+0.040}_{-0.039}   $\\
$\omega_{cdm}$  		 &\ \ $0.1183^{+0.0030}_{-0.0031}$	 & \ \ $0.1203^{+0.0032}_{-0.0031}$	&\ \ $0.1198^{+0.0030}_{-0.0031}$	&\ \ $0.1198^{+0.0031}_{-0.0030}$	\\
$H_{0}$ 				 &\ \ $67.6^{+1.6}_{-1.8}        $ 	 & \ \ $71.6^{+2.1}_{-2.1}        $	&\ \ $72.4^{+2.0}_{-2.0}        $	&\ \ $72.3^{+2.1}_{-2.1}        $\\
$10^{+9}A_{s}$  		 &\ \ $2.16^{+0.13}_{-0.12}      $ 	 & \ \ $2.10^{+0.14}_{-0.14}      $	&\ \ $2.10^{+0.14}_{-0.13}      $	&\ \ $2.09^{+0.14}_{-0.14}      $\\
$n_s$  					 &\ \ $0.9649^{+0.0099}_{-0.0099}$	 & \ \ $0.9604^{+0.0097}_{-0.0096}$	&\ \ $0.9607^{+0.0097}_{-0.0091}$	&\ \ $0.9607^{+0.0096}_{-0.0097}$\\
$\tau_{reio}$ 			 &\ \ $0.073^{+0.033}_{-0.031}   $ 	 & \ \ $0.056^{+0.035}_{-0.037}   $	&\ \ $0.056^{+0.035}_{-0.035}   $	&\ \ $0.055^{+0.035}_{-0.037}   $\\
$\sum m_\nu\ [{\rm eV}]$ &\ \ $< 0.351 (2\sigma)$ 	 		 & \ \ $0.56^{+0.21}_{-0.19}      $	&\ \ $0.51^{+0.19}_{-0.19}      $	&\ \ $0.51^{+0.21}_{-0.19}      $\\
\\
\hline
\\
$\Omega_{smg}$  		 &\ \ $--$ 		  					 & \ \ $0.710^{+0.021}_{-0.023}   $	&\ \ $0.718^{+0.020}_{-0.021}   $	&\ \ $0.718^{+0.020}_{-0.022}   $		\\
$\xi$  					 &\ \ $--$ 		  					 & \ \ $2.064^{+0.031}_{-0.033}   $	&\ \ $2.41^{+0.20}_{-0.20}      $	&\ \ $2.39^{+0.61}_{-0.68}      $	\\
$c_2$  					 &\ \ $--$ 		  					 & \ \ \fix{-1}						&\ \ \fix{-1}						&\ \ \fix{-1}							\\
$c_3$  					 &\ \ $--$ 		  					 & \ \ $-0.0807^{+0.0012}_{-0.0013}$	&\ \ $-0.1042^{+0.0096}_{-0.0077}$	&\ \ $-0.074^{+0.27}_{-0.077}   $	\\
$c_4$  					 &\ \ $--$ 		  					 & \ \ \fix{0}			 			&\ \ $-0.0048^{+0.0018}_{-0.0014}$	&\ \ $0.008^{+0.11}_{-0.026}    $ 		\\
$c_5$  					 &\ \ $--$ 		  					 & \ \ \fix{0}			 			&\ \ \fix{0}						&\ \ $-0.013^{+0.023}_{-0.12}   $		\\

\\
\hline \\
$\chi^2_{\rm CMB}$						 &\ \ 11,273.7	  & \ \ 11,288.0				 			&\ \ 11,275.1				&\ \  11,274.0 \\
$\chi^2_{\rm BAO13}$				 		 &\ \ 1.75   	  & \ \ 0.90  								&\ \ 0.83 				&\ \  0.87 	 \\ \\
\hline \hline
\end{tabular}
\caption{One-dimensional marginalized CMB+BAO13 constraints at 95\% confidence level on the parameters of the cubic, quartic and quintic Galileon models studied in this paper, together with $\lcdm$. The constraints correspond to a dataset that comprises temperature and lensing data from Planck, as well as BAO13 constraints. Bold values indicate values that are fixed, i.e., not varied in the MCMC analysis. The last lines indicates the goodness-of-fit to CMB and BAO13 data separately.}
\label{table:params}
\end{table*}
\begin{figure}
 \centering
 \begin{minipage}{0.49\linewidth}
    \centering
   {\includegraphics[width=\textwidth]
   {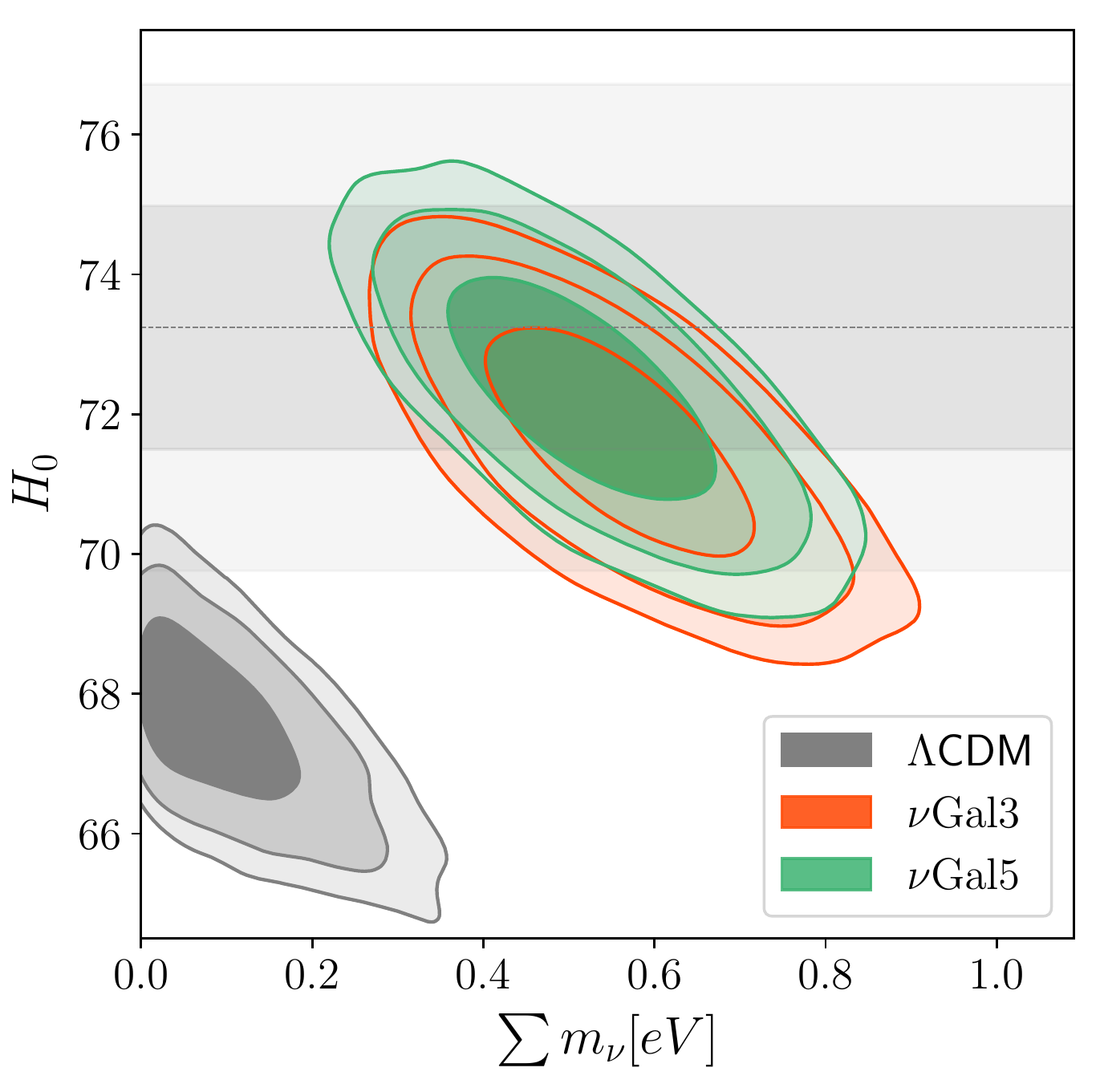}} \\
 \end{minipage}
 \hfill
\begin{minipage}{0.49\linewidth}
    \centering
	{\includegraphics[width=\textwidth]
	{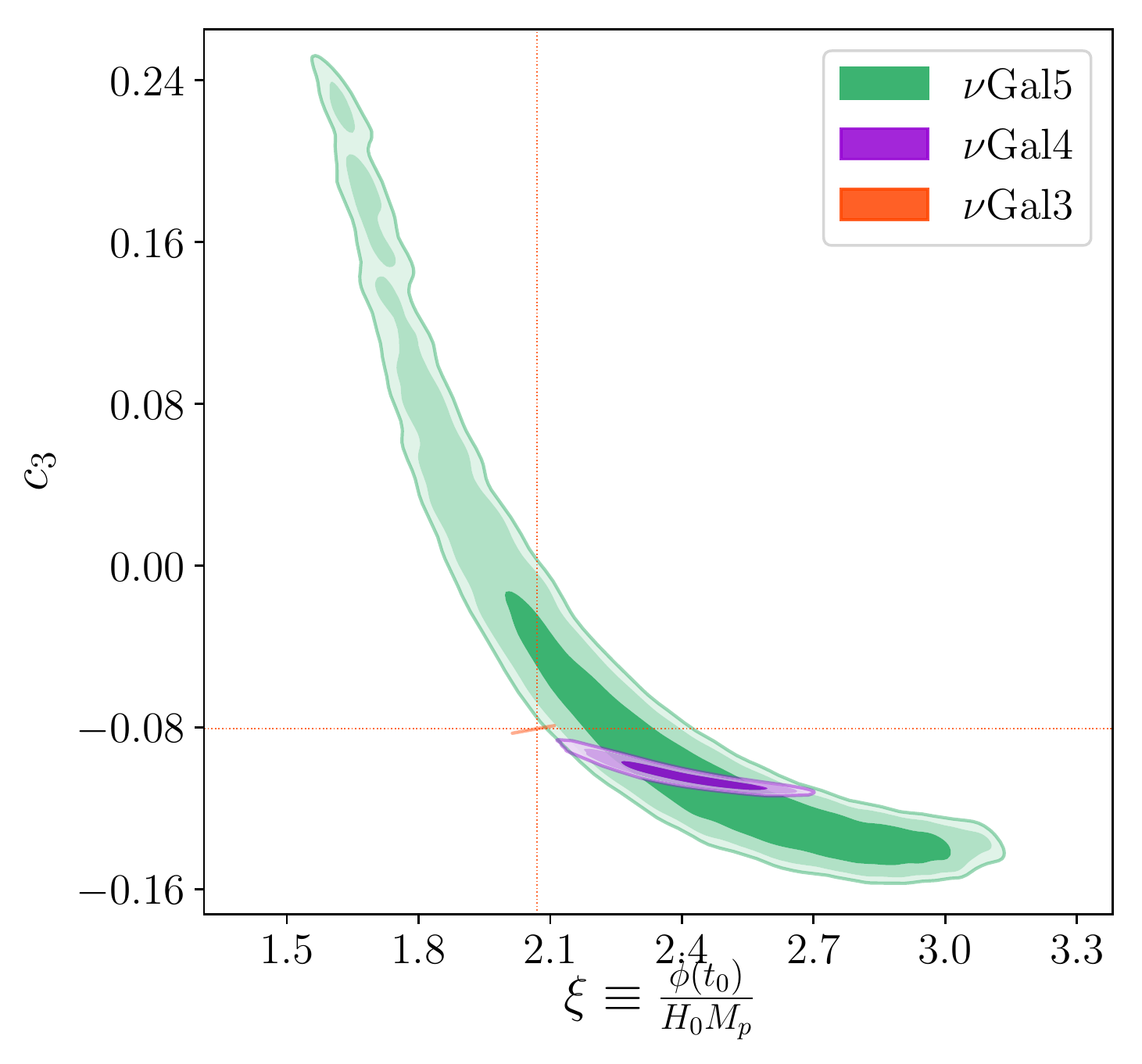}} \\
\end{minipage}

\caption{1-3 $\sigma$ contours in the 2-D marginalized $H_0 - \Sigma m_\nu$ plane (\lp) and in the $\xi - c_3$ plane (\rp) from the MCMCs with CMB and BAO13 data for $\lcdm$, Cubic, Quartic and Quintic Galileons. The horizontal shaded regions in the {\lp} indicate the constrains on $H_0$ from local (distance ladder) measurements \cite{Riess:2016jrr}. We have omitted the quartic model in the $H_0 - \Sigma m_\nu$ plane for the purpose of clearness since the constrains are almost indistinguishable from the quintic case. The red, dotted lines in the {\rp} point to the contours of Cubic Galileons.}
\label{fig:xic3-H0}
\end{figure}

\subsection{Monte Carlo Cosmological Constraints}
The one-dimensional marginalized constraints obtained with the CMB+BAO13 dataset for the Galileon models are listed in \autoref{table:params}. In agreement with \cite{Barreira:2014jha}, we find that the constraints on the cosmological parameters (cf. Eq.~(\ref{eq:params})) are practically the same across the cubic, quartic and quintic models. This is because these constraints are largely set by $H(a)$, which on the tracker, is independent of the values of the Galileon parameters $c_i, \xi$ and hence it is the same for all three sectors of the model. \autoref{table:params} shows also the corresponding results for $\lcdm$ to help appreciate the difference in the resulting best-fitting parameter values. Two noteworthy such differences are those associated with the constraints on $H_0$ and $\Sigma m_\nu$, as illustrated in the \lp\ of \autoref{fig:xic3-H0} and on which we comment next.

In Galileon gravity cosmologies the data require non-zero neutrino mass values which is in sharp contrast with the result in $\lcdm$ \cite{Ade:2015xua,Palanque-Delabrouille:2015pga}. Specifically, in Galileon cosmologies the data rule out $\Sigma m_\nu \neq 0$ with $\approx 5\sigma$ significance; while in $\lcdm$ $\Sigma m_\nu = 0$ is favoured by the data. As explained first in \cite{Barreira:2014jha}, given the tracker expansion rate of Eq.~(\ref{eq:hubletracker}), high neutrino mass values are \emph{needed} for the model to simultaneously fit the peak positions of the CMB at early redshift and BAO features at low redshift. These constraints on $\Sigma m_\nu$ open the route for cosmological-independent terrestrial determinations of the absolute neutrino mass scale to help distinguish between $\lcdm$ and Galileon cosmologies. Currently, these efforts are limited to a sensitivity of $\sum m_{\nu} < 6$ eV \cite{Drexlin:2013lha}, but future experiments are expected to improve this significantly.

The value of $H_0$ preferred by the CMB+BAO13 dataset in the constraints of the Galileon model is in agreement with the measurement in the local Universe reported in \cite{Riess:2016jrr} (see also \cite{Cardona:2016ems,Bonvin:2016crt,Feeney:2017sgx,Zhang:2017aqn,Follin:2017ljs,Dhawan:2017ywl,Wu:2017fpr}). This agreement occurs without adding any prior on $H_0$. As for $\Sigma m_\nu$, the different constraints on $H_0$ can be traced back to the details of the evolution of $H(a)$ in the $\lcdm$ and Galileon models. This difference to $\lcdm$ gains particular relevance when interpreted in light of the current 3.4$\sigma$ tension in $\lcdm$ between the CMB inferred value of $H_0$ and the local determination which has been the subject of recent investigation \cite{2016arXiv160705677L, 2016arXiv160705617B}.

The \rp\ of \autoref{fig:xic3-H0} shows the two-dimensional marginalized constraints on the $c_3$-$\xi$ plane. Note that these two parameters are only independent in the quintic case (cf.~\autoref{sec:gal-bg}); the contours of the $\nu{Gal}3$ model are barely visible at $\xi \approx 2.1$ and $c_3 \approx -0.08$. We note also that, in addition to the constraints from the CMB+BAO13 dataset, the parameter space of the quintic Galileon model is also severely constrained by the stability conditions (cf. Figure 12 of \cite{Barreira:2014jha}).

Our goal in this paper is not to undergo a detailed analysis of these cosmological constraint results. Instead, we limit ourselves to noting that \autoref{fig:xic3-H0} serves as a useful reminder that cosmological parameter constraints are model-dependent in general and that some of the observational tensions that have been reported in $\lcdm$ may be circumvented by alternative theoretical models. Below, we analyse with more detail the ISW predictions of these best-fitting regions and what they imply for the viability of the Galileon model.

\subsection{Serious Tension of the Cubic Galileon and the ISW Data}
Overall, we find that the phenomenology of the Cubic Galileon is not flexible enough to fit the ISW data. More specifically, sampling from the points accepted in the MCMC analysis we found no single point with a positive value of $\Aisw$, i.e., the lensing potential in these cubic Galileon models always grows during the redshift range covered by the WISE galaxies. The model with the smallest tension ($\xi =2.04$, $c_3 = -0.08$) has an ISW amplitude of $\Aisw = -2.39$. This prediction is in a $7.8 \sigma$ tension with the best fit amplitude (Eqs.~(\ref{eq:Abf}) \& (\ref{eq:sigmabf})) to the WISE ISW signal. The orange dashed line in \autoref{fig:Pot-ISW} shows the ISW signal for this poor best-fitting case (with galaxy bias as in \autoref{fig:kappa-g_zdep}).

Given this very strong and apparently unavoidable tension with the ISW data we can conclude that the covariant Cubic Galileon is not a viable cosmological model. In \cite{kimura2} the authors have also reached similar conclusions for a model of gravity that has the Cubic Galileon as a specific limit; our analysis is however more robust as (i) we analyse the regions of the parameter space that best fit the CMB and BAO13 data, which include having non-zero neutrino masses; and (ii) we perform the calibration of the bias of the WISE galaxies (cf.~\autoref{sec:bias}), which, if not done, constitutes a source of error on the overall ISW signal prediction.

%
%
\subsection{ISW Constraints on the Quartic and Quintic Galileon Models}
\begin{figure}
 \centering
  \begin{minipage}{0.80\linewidth}
    \centering
   {\includegraphics[width=\textwidth]
   {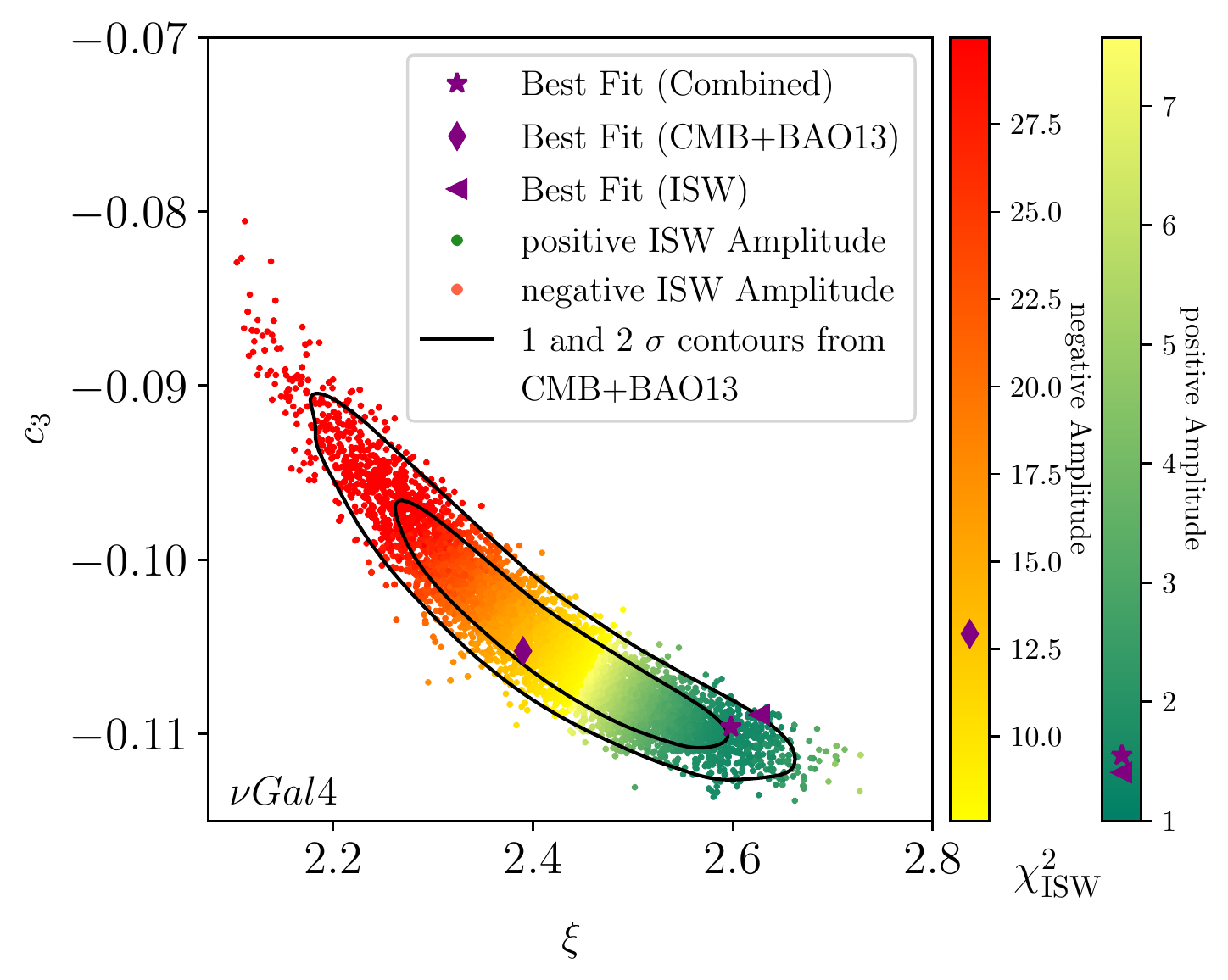}} \\
  \end{minipage}

  \caption{ISW amplitude $\Aisw$ across the $c_3-\xi$ plane of the Quartic Galileon parameter space. The solid contours denote the two-dimensional marginalized $1\sigma$ and $2\sigma$ confidence regions from the MCMCs with CMB+BAO13 data. The dots correspond to points accepted in the MCMCs and are colour coded by their corresponding $\chi^2_{\rm ISW}$ values. We use different colourbars for points with positive and negative $\Aisw$ to facilitate interpreting the figure. Note that all models leading to a $\chi^2_{\rm ISW} > 30$ are shown in dark red. The rhombus, triangle and star symbols in purple indicate the models that give the best fit to the CMB+BAO13 dataset, ISW data alone and the combined CMB+BAO13+ISW set, respectively.}
  \label{fig:ISW_Gnu4}
\end{figure}

\begin{figure}
 \centering
  \begin{minipage}{0.80\linewidth}
    \centering
   {\includegraphics[width=\textwidth]
   {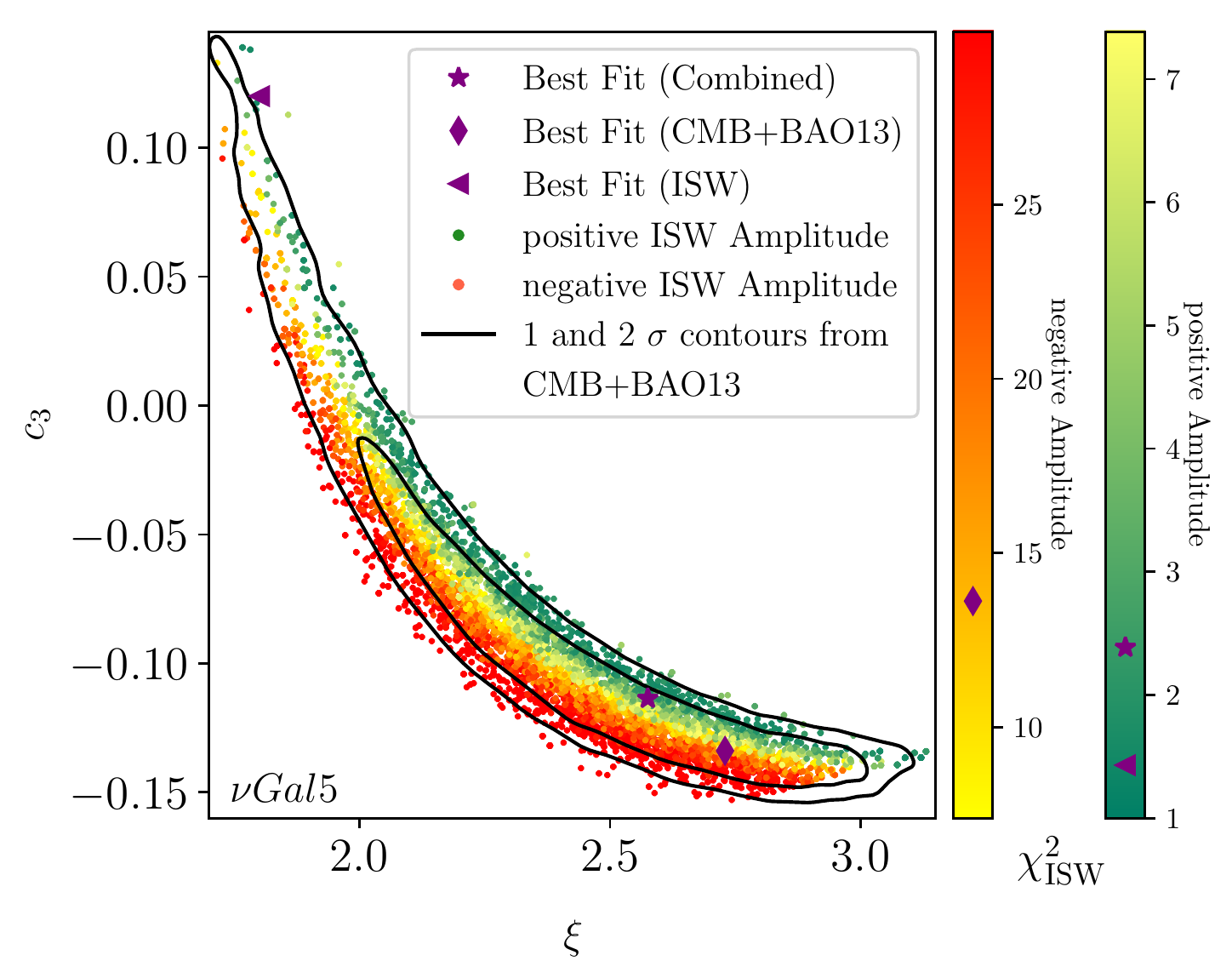}} \\
  \end{minipage}

   \caption{Same as \autoref{fig:ISW_Gnu4} but for the quintic Galileon model. See the main text for comments about the sampling at the tips of the contours.}
 
  \label{fig:ISW_Gnu5}
\end{figure}

\begin{table}
\begin{adjustbox}{center}
\begin{tabular}{l|ccccccccccc}
& $100\omega_b$ & $\omega_{cdm}$ & $\Omega_{smg}$ & $H_{0}$ & $10^{+9}A_{s}$ & $n_s$ & $\tau_{reio}$ & $\sum m_\nu$& $\xi$& $c_3$ \\ \hline
$\nu Gal 4$ & 2.193 & 0.1195 & 0.7264 & 73.08 & 2.04 & 0.957 & 0.0439 & 0.44 & 2.60 & -0.11 \\
$\nu Gal 5$ & 2.204 & 0.1193 & 0.7093 & 71.39 & 2.13 & 0.963 & 0.0647 & 0.63 & 2.58 & -0.11

\end{tabular}
\end{adjustbox}
\caption{Best-fitting cosmological and Galileon parameters to CMB, BAO13 and ISW data sets for Quartic and Quintic Galileons, with $\chi^2_{\rm CMB} = 11,277.9$, $\chi^2_{\rm BAO13} = 1.85$, $\chi^2_{\rm ISW} = 1.55$ and $\chi^2_{\rm CMB} = 11,276.6$, $\chi^2_{\rm BAO13} = 0.98$, $\chi^2_{\rm ISW} = 2.39$, respectively. Both models are within the $\approx 1 \sigma$ confidence region of the MCMC constraints.}
\label{tab:CombBF}
\end{table}

In the quartic and quintic Galileon models the additional terms introduced in the Lagrangian and extra degrees of freedom ($\xi$ in the quartic model and $\xi, c_3$ in the quintic; cf.~\autoref{sec:gal-bg}) allow for a time evolution of the lensing potential that results in positive $\Aisw$ values in the redshift range of the WISE survey (see \cite{DeFelice:2011aa} for a discussion about the importance of $c4,c5 \neq 0$ in the sign of the ISW effect.). In \autoref{fig:ISW_Gnu4} and \autoref{fig:ISW_Gnu5} we show the two-dimensional marginalized constraints in the $c_3$-$\xi$ plane of the quartic and quintic models. These two figures show also some accepted MCMC points colour coded by the respective $\chi^2_{\rm ISW}$ values; we use two colourbars for points with $\Aisw>0$ and points with $\Aisw<0$ to visualize which regions of the parameter space predict a positive/negative ISW signal. The purple symbols mark the location of the best-fitting model to the CMB+BAO13 data only (\emph{rhombus}), ISW data only (\emph{triangle}) and to the combined data sets (\emph{star}). By summing the $\chi^2$ values of CMB, BAO13 and ISW (assuming that the likelihoods are independent) we obtain the global best fits shown in \autoref{tab:CombBF}.

For both, the quartic and quintic models, there are parameter space regions that are good fits to simultaneously the CMB+BAO13 dataset and the $C_{\ell}^{\rm{Tg}}$ spectra of the WISE galaxies: the best-fitting models to this combined data are in fact within the $\approx 1\sigma$ limits of the original CMB+BAO13 constraints. The $C_{\ell}^{\rm{Tg}}$ spectra and lensing potentials of the best-fitting models to the CMB+BAO13 dataset (purple rhombus) and to its combination with ISW (purple stars) are shown by the dashed and solid lines in \autoref{fig:Pot-ISW}, respectively. The corresponding best-fitting galaxy bias values for the latter models are given in \autoref{fig:kappa-g_zdep}.

As a technical remark, we note that the low-$\xi$/high-$c_3$ tail of the distribution in \autoref{fig:ISW_Gnu5} is hard to sample by standard MCMC algorithms because it is very narrow and the neighbouring regions correspond to points associated with ghost and Laplace instabilities (whenever these unstable points are sampled they are immediately rejected before even obtaining predictions for them). In the other tip of the contours (high-$\xi$/low-$c_3$), the parameter space is also sharply cut off by the stability conditions; in fact, the algorithm that determines the confidence contours cannot resolve these fine details of the Galileon parameter space. For us, the main point to retain is that there are parts of the Galileon subspace of parameters that yield a good fit to the ISW signal of the WISE galaxies.

\begin{figure}
 \centering
  \begin{minipage}{0.49\linewidth}
    \centering
   {\includegraphics[width=\textwidth]
	{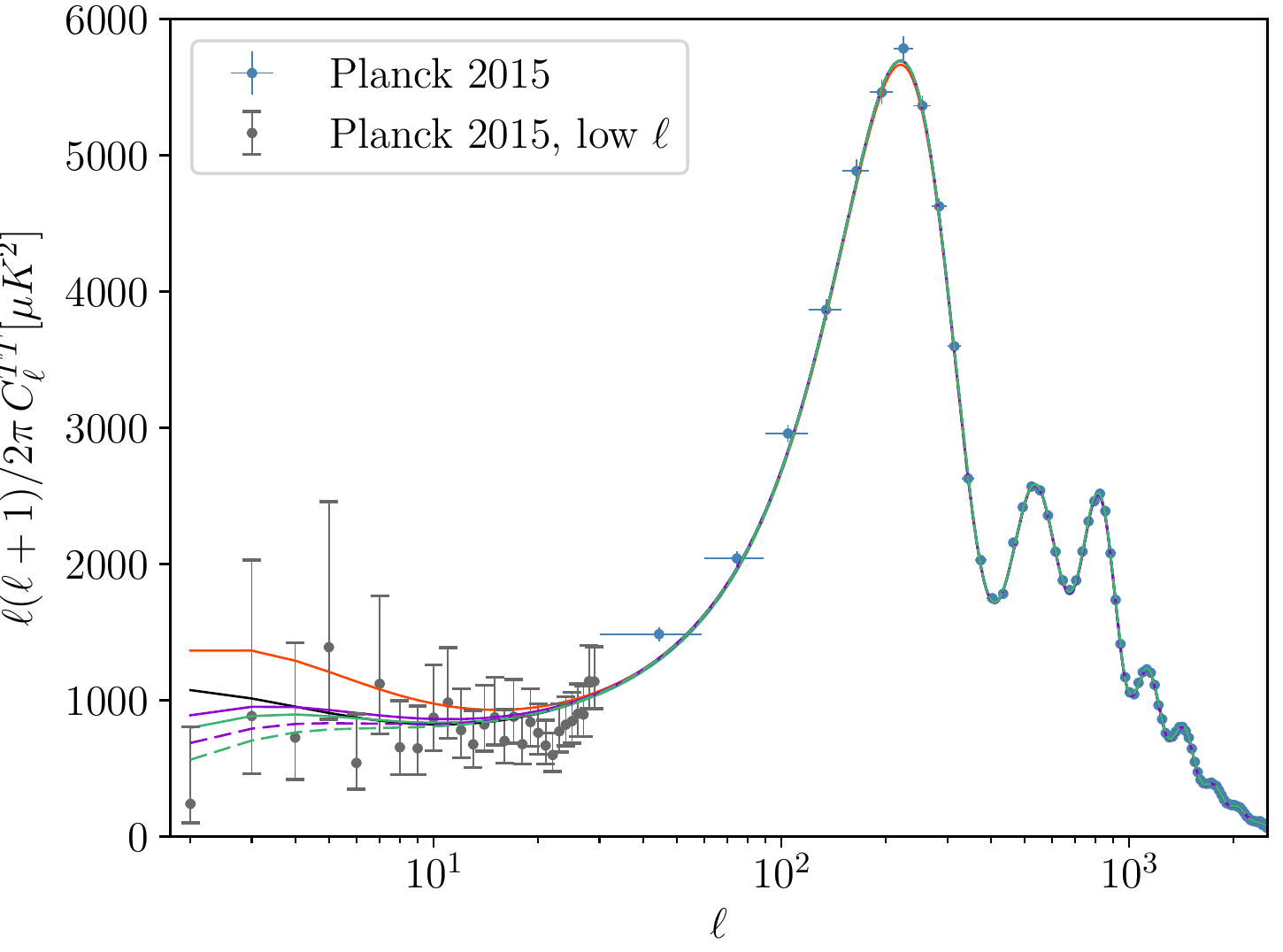}} \\
  \end{minipage}
   \hfill
   \begin{minipage}{0.49\linewidth}
    \centering
	{\includegraphics[width=\textwidth]
    {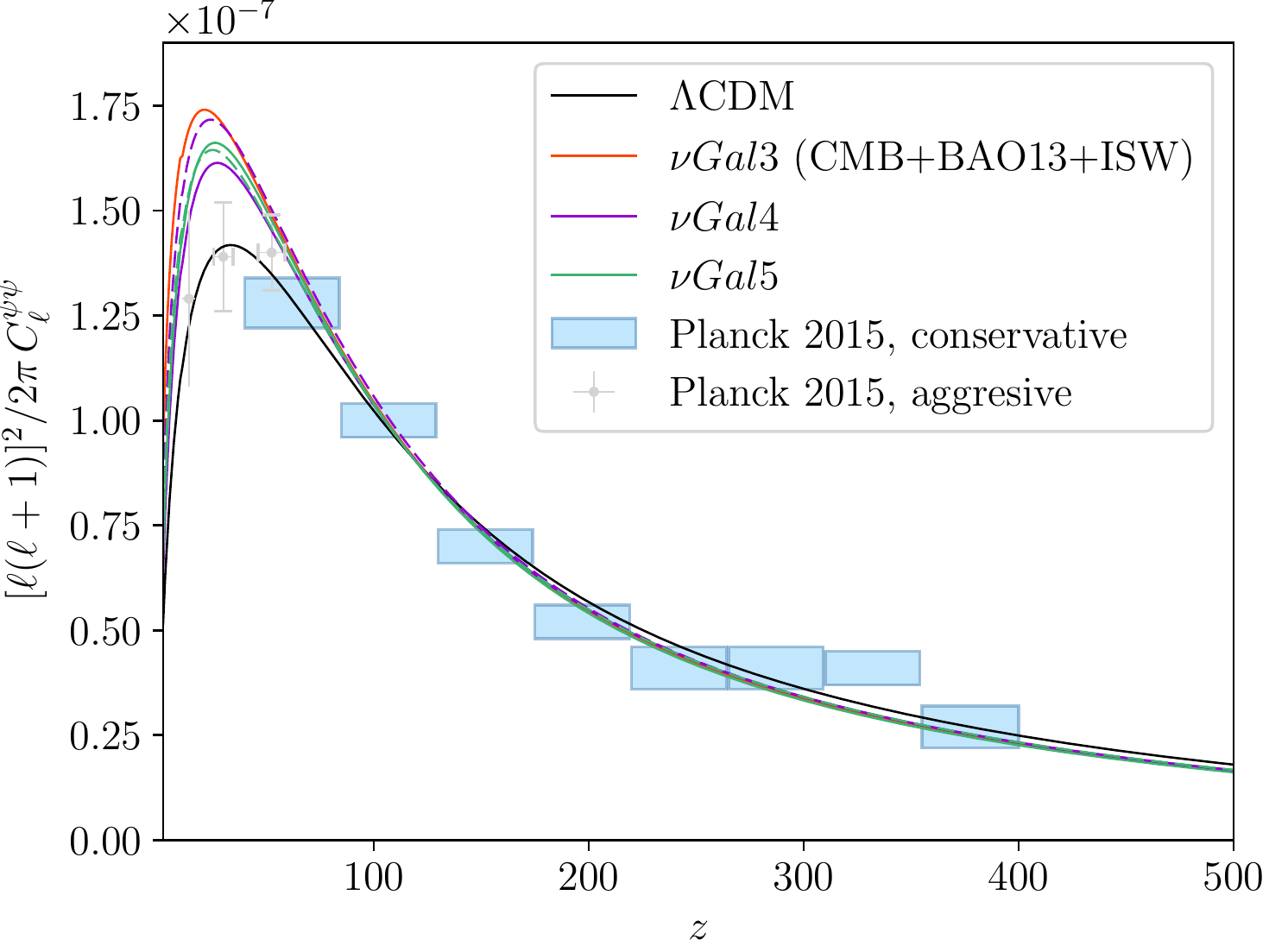}} \\
  \end{minipage}
  \caption{CMB temperature (\lp) and lensing potential (\rp) power spectrum for $\Lambda$CDM and the Galileon models. Dashed lines correspond to Galileon models that provide the best fit to CMB+BAO13 data only, while solid lines are the models with the best fit to CMB+BAO13 data combined with measurements of the ISW effect obtained with the WISE galaxy survey. Error bars from Planck 2015 are indicated in grey/blue. In the {\rp} the blue, shades regions correspond to the lensing potential obtained with ``conservative'' binning while the grey bars show the low $\ell$ data from the ``aggressive'' binning method from Table 1 of \cite{Ade:2015zua}. Note that the latter points are not included in the Planck likelihood.} 
  \label{fig:CMB-TT-pp}
\end{figure}

The CMB temperature and lensing potential power spectrum of the best-fitting Galileon models to the CMB+BAO13 data set (dashed) and its combination with ISW data (solid) are shown in \autoref{fig:CMB-TT-pp}, together with the data points and errorbars from the Planck 2015 data release \cite{Ade:2013sjv,Ade:2015zua}, as labelled. The figure illustrates the overall good fit of the Galileon models that survive the WISE ISW test. In the right panel, the two data symbols shown correspond to a conservative and a more aggressive treatment of the power spectrum of the reconstructed lensing potential maps. In the official Planck lensing likelihood only the conservative points are included and as a result, the Galileon models and $\lcdm$ display similar goodness-of-fit. It is interesting to note that, compared to $\lcdm$, the Galileon models predict a markedly larger amplitude at low-$\ell$. However, systematics on the determination of the data points at low$-\ell$ are currently less well understood compared to high-$\ell$. Hence, including these points in a constraint analysis could lead to potentially biased results (as a matter of fact, CMB lensing data are not included in the analysis of the Planck paper dedicated to dark energy and modified gravity \cite{Ade:2015rim}). For the time being we limit ourselves to noting that a more robust understanding of the CMB lensing potential power spectrum at low-$\ell$ could prove very useful in distinguishing between $\lcdm$ and Galileon gravity.

As noted already above, the positiveness of the WISE ISW signal in the Quartic and Quintic Galileons follows directly from the fact that the lensing potential in these models is decaying during the redshift range covered by the WISE galaxies, as illustrated in the \lp \ of \autoref{fig:Pot-ISW}. The same panel also shows, however, that the lensing potential can grow at other epochs: $z\sim0.5-1$ for the best-fitting quartic and quintic cases shown there. A prediction of these models is therefore that the sign of $\Aisw$ is in general a function of redshift; this is different than in $\lcdm$ in which the potentials always decay after the onset of the acceleration of the Universe. An interesting observational test to perform is therefore that of computing $\Aisw$ at a sufficiently fine series of redshift bins $\Delta z$ by measuring $C_{\ell}^{{\rm Tg}}$ using galaxy samples that cover those same redshift bins. To the best of our knowledge, there are currently no such measurements to readily perform such a test. As a check, we have computed the prediction of the quartic and quintic Galileon models for the cross-correlation of CMB temperature with the galaxy distribution of the NRAO VLA Sky Survey (NVSS) \cite{1998AJ....115.1693C}, as well as with the CMB lensing maps. These two probes are sensitive to the time-evolution of the potentials during a much wider redshift range compared to the WISE galaxy sample. The NVSS selection function peaks around $z \sim 0.3$ and spreads out to redshift $z \sim 5$, while with CMB lensing one is sensitive out to the redshift of recombination via the lensing kernel (see Fig.~3 of \cite{Ade:2015dva} for an illustration). We find that the predicted signals for the best-fitting models from \autoref{tab:CombBF} are within the corresponding $1\sigma$ bounds reported in the Planck paper \cite{Ade:2015dva} (upper left panel for NVSS and lower right panel for lensing in Figure 6 there).

\subsection{Tension with BAO Data}
\label{sec:BAO-tens}
\begin{figure}
 \centering
  \begin{minipage}{0.75\linewidth}
    \centering
   {\includegraphics[width=\textwidth]
   {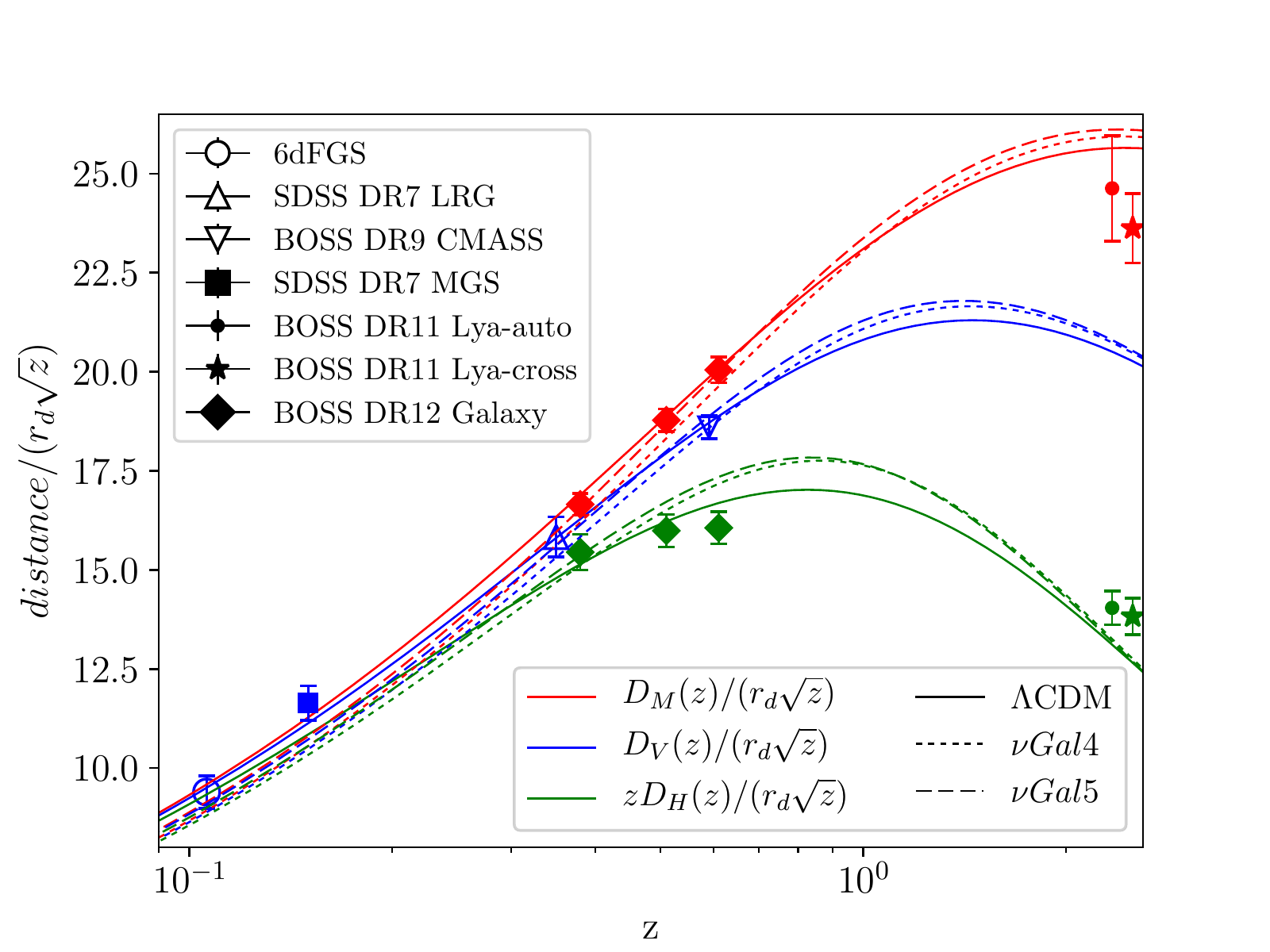}} \\
  \end{minipage}
  \caption{Redshift dependence of the $D_M$, $D_V$ and $D_H$ distance scales for $\lcdm$ 
   and for the best-fitting $\nu Gal4$ and $\nu Gal5$ models to the CMB+BAO13+ISW data, as labelled. The symbols with errorbars represent the determinations from the surveys indicated in the legend: 6dFGS \cite{beutler20116df}, SDSS DR7 LRG \cite{padmanabhan20122}, BOSS DR9 CMASS \cite{anderson2012clustering}, SDSS DR7 MGS \cite{Ross:2014qpa}, BOSS DR11 Ly$\alpha$-auto \cite{2015A&A...574A..59D}, BOSS DR11 Ly$\alpha$-cross \cite{Font-Ribera:2013wce} and BOSS DR12 Galaxy \cite{Alam:2016hwk}. The symbols left unfilled comprise the BAO13 dataset used in our MCMCs. The distances are divided by the sound horizon (computed whit \hiclass) at the end drag epoch, $r_d = 147.31$ Mpc, $r_d = 147.16$ Mpc and $r_d = 147.44$ Mpc for $\lcdm$, $\nu Gal4$ and $\nu Gal5$, respectively. The scaling by $z, \sqrt{z}$ serve to bring all curves to a similar dynamical range in the y-axis.
   }
   \label{fig:baodist}
\end{figure}
\begin{table}
\centering
	\begin{tabular}{l|cccc}
	& $\chi^2_{\rm MGS}$ & $\chi^2_{\rm DR12}$ & $\chi^2_{\rm Ly\alpha-auto}$ &  $\chi^2_{\rm Ly\alpha-cross}$ \\ \hline 
data points & 1    & 6 		& 2    & 2 		\\ \hline \hline
$\lcdm$ 	& 1.98 & 5.02   & 4.95 & 4.78  	\\
$\nu Gal 4$ & 5.82 & 12.69  & 5.35 & 3.61 	\\
$\nu Gal 5$ & 3.45 & 12.90  & 6.05 & 4.26 
\end{tabular}
\caption{Goodness-of-fit to the BAO measurements not included in the MCMCs for $\lcdm$ and for the best-fitting (to CMB+BAO13+ISW data) quartic and quintic Galileon model from \autoref{tab:CombBF}.} \label{tab:BAO_chi2}
\end{table}

The redshift dependence of three distance scales constrained by BAO analyses is shown in \autoref{fig:baodist}: $D_M(z) = (1+z)D_A$, $D_H(z) = c/H(z)$ and $D_V(z) = \left[zD_H(z)D_M(z)^2\right]$, where $D_A(z)$ is the physical angular diameter distance. The result is shown for $\lcdm$ and for the best-fitting Quartic and Quintic Galileons to the CMB+BAO13+ISW data, as labelled. The symbols with errorbars display the determinations obtained by various BAO analyses. The BAO compilation that we used in our MCMC analysis is marked with unfilled symbols and, as already noted, both $\lcdm$ and the Galileon models provide good fits to these data.

The determinations of the BAO distance scales that are more recent than those in our BAO13 compilation are marked by the filled symbols. As we have anticipated before when leaving these data out of our MCMCs, the figure shows that the predictions of the best-fitting Galileons are in tension with these data. The $\chi^2$ values listed in \autoref{tab:BAO_chi2} illustrate this more quantitatively. The strongest of the tensions is with the SDSS MGS value \cite{Ross:2014qpa}: the $\nu Gal4$ and $\nu Gal5$ are $\approx 2.4\sigma$ and $\approx 1.9\sigma$ away from this measurement, respectively. For the case of the BOSS DR12 points (BAO-only column in table 7 of \cite{Alam:2016hwk}), the \emph{tension} is approximately at the $1.5\sigma$ level for both $\nu Gal4$ and $\nu Gal5$ (estimated as $\sqrt{\chi^2/dof}$; note also that the 6 data points are correlated, which makes this only a rough estimate). Although at face value, $1.5\sigma$ is not a significant tension for the fit to the 6 BOSS DR12 data points, we note that the $\chi^2 \approx 13$ values are dominated by the higher-$z$ $D_H$ points, which the Galileon model fits poorly. This is why we dub this as a tension. The BAO determinations from analyses of the Ly$\alpha$ forest of BOSS quasars \cite{2015A&A...574A..59D, Font-Ribera:2013wce} are in tension with Galileon model as well as with $\lcdm$. Here, our results suggest that if the current tension in $\lcdm$ persists in future higher-fidelity analysis, then the tension is unlikely to be resolved by Galileon-like modifications to gravity alone.

The $\chi^2$ values quoted in \autoref{tab:BAO_chi2} and the curves shown in \autoref{fig:baodist} correspond to the specific case of the best-fitting quartic and quintic models to the CMB+BAO13+ISW dataset. We have explicitly checked nonetheless that the degree of tension for DR12 and Lyman $\alpha$ data is representative of all points within the $2\sigma$ contours obtained with the MCMC with CMB+BAO13 data; although we note that in the quintic case the tension to the MGS data point can be relieved ($\chi^2_{\rm MGS} < 1.5$) within the ISW-compatible $1\sigma$ contours from the MCMCs.

Appreciable as the tensions identified above are, they do not yet allow us to confidently conclude that they rule out the Galileon cosmologies. The different BAO distance scales in LCDM and Galileon gravity depicted in \autoref{fig:baodist} do assign, however, to future BAO data great potential to distinguish between these cosmological models.

For completeness we apply a simple model selection criterion that takes into account the extra degrees of freedom of the Galileon models compared to $\lcdm$. We use the Akaike Information Criterion (AIC)%
\footnote{AIC$=2k + \Sigma_i \chi^2_i$, with the number of model parameters k and $\chi^2_i$ being the $\chi^2$ values for the different, independent data sets $i$. The relative likelihood -- quantifying by which factor a model is more likely than the fiducial model -- is given by ${\rm exp((AIC_{fid}- AIC_{m})}/2)$.} %
\cite{akaike1974new} to compute the relative likelihood for the best-fitting quartic and quintic Galileon model w.r.t. to $\lcdm$ for different combinations of data sets, which we assume to be independent%
\footnote{Strictly speaking the assumption of independence it not true when considering all BAO data points: the BOSS CMASS sample enters not only from DR9 but is also included in the DR12 Galaxy sample. However, we neglect this correlation in this rough estimate as it just serves to build an intuition about the ``cost'' of introducing one/two more parameters w.r.t. $\lcdm$.}. %
Here, we also consider the local $H_0$ measurement from \cite{Riess:2016jrr}, where $H_0 = 73.24 \pm 1.74$. The results are shown in \autoref{tab:ModelComp}.

\begin{table}
\begin{adjustbox}{center}
\begin{tabular}{l|c|c}
Data Sets & $\nu Gal 4$ &  $\nu Gal 5$ \\ \hline \hline
CMB+BAO13+ISW 				& 1/25					& 3/100					 \\ 
CMB+BAO13+ISW+$H_0$			& 12/1					& 5/1					 \\ \hline
CMB+BAO17+ISW 				& 1/5000				& 1/5000				 \\ 
CMB+BAO17+ISW+$H_0$			& 3/50					& 1/25					 \\ 

\end{tabular}
\end{adjustbox}
\caption{Relative likelihood of the best-fitting quartic and quintic Galileon model compared to $\lcdm$ for a combination of different data sets computed with the Akaike Information Criterion; the relative likelihood is given by ${\rm exp((AIC_{\lcdm}- AIC_{Gal})}/2)$, indicating by which factor a model is more or less likely than $\lcdm$. BAO17 includes all BAO measurements from \autoref{fig:baodist}. According to the AIC Galileons are favoured when considering the local $H_0$ measurement, but not when simultaneously taking into account new BAO data. The respective values for the cubic models are all disfavoured by a factor of more than $1/10^{16}$ due to the strong tension with the WISE ISW data. As a rule of thumb \cite{2007MNRAS.377L..74L}, if these values are smaller than $\approx 1/13$ and $\approx 1/150$, then this constitutes ``strong'' and ``decisive'' preference by the data for $\lcdm$ over the Galileon model, respectively.}
\label{tab:ModelComp}
\end{table}

Owing to the extra degrees of freedom, Galileon models are disfavoured when considering only CMB, BAO and ISW data. It is interesting to note however, that the inclusion of the local measurement of the Hubble constant from \cite{Riess:2016jrr} eases this disfavour%
\footnote{If local measurements of the Hubble constant with larger errorbars are considered, as e.g. $H_0 = 70.6 \pm 3.3$ km/s/Mpc from \cite{Efstathiou:2013via}, the Galileon models will keep being disfavoured due to their extra degrees of freedom. The respective values for the quartic case are 1/18 (BAO13) and 1/4300 (BAO17). For the quintic case one obtains 1/19 (BAO13) and 1/3000 (BAO17).}. %
More robust model comparison analyses can be performed by computing the actual Bayesian evidence \cite{Trotta:2008qt} of the competing models. Such a dedicated model comparison analysis would also benefit from the use of a wider dataset than that considered in this paper (adding to it for example lensing shear, growth rate, additional ISW data; see below).

\subsection{On Additional Datasets: SNIa and Growth Rate Data}
In this paper, we did not explicitly include constraints from type Ia supernovae (SNIa). This was motivated to reproduce/corroborate the constraint analysis of \cite{Barreira:2014jha} who also do not show results from SNIa constraints. To validate that there are not tensions arising from also considering SNIa data we carried out additional MCMC analyses for the Galileon models with CMB data and 740 SNIa supernova from the ``Joint Light-curve Analysis'' (JLA) data set \cite{betoule2014improved}. We found in this check: (i) BAO data have more constraining power than SNIa and (ii) we found no tension in the Galileon models between the two data sets.

Measurements of the growth rate of structure $f = {\rm dln}\delta/{\rm dln}a$ (where $\delta$ is the linear density contrast) could also play an important role in the constraints of the Galileon model. These estimates of the growth rate, which is normally quoted as the combination $f\sigma_8$, are obtained from galaxy clustering data by fitting to it a model of RSD, galaxy bias and non-linear clustering (RSD-bias-nonlinear model; see e.g.~\cite{2016arXiv160703147S} for an example of such analysis from the BOSS survey). Modified gravity theories can then only be constrained with these $f\sigma_8$ estimates if they are compatible with the assumptions that go into the RSD-bias-nonlinear model. Normally, these latter modelling steps are based on GR (see \cite{2014PhRvD..89d3509T, 2016JCAP...08..032B, 2017arXiv170202348B} for recent exceptions to this), and as a result their performance on other theories of gravity is not guaranteed to be unbiased. The standard way to test these RSD-bias-nonlinear models is to build mock catalogues based on N-body simulations of the various theories of gravity and check whether the model recovers the $f\sigma_8$ value of the input cosmology. Such a recent analysis was carried out in \cite{2016PhRvD..94h4022B} for the normal branch of the Dvali-Gabadadze-Porrati (DGP) \cite{Dvali:2000hr} model using the RSD-bias-nonlinear used in \cite{2016arXiv160703147S}. In the case of the Galileon model extra complications arise because of the non-negligible scale-dependency induced on the growth rate from the large fraction of massive neutrinos. This is also not normally taken into account in the observational determinations of $f\sigma_8$ and thus, prevents us from using the current data to constraint Galileon gravity further.

%
%

\section{Summary and Conclusion} \label{sec:conc}
We have carried out an investigation of the observational viability of cosmologies with covariant Galileon gravity as alternatives to standard $\lcdm$ using CMB, BAO and ISW data. In the Galileon model the departures from standard GR are controlled by a scalar field whose couplings to the metric field (i) modify the gravitational force law and hence leave signatures on observables sensitive to structure formation in the Universe; and (ii) naturally yield self-accelerating background solutions, i.e., can explain the observed acceleration of the expansion of the Universe without a cosmological constant. The Galileon model does not have a $\lcdm$ limit and has a rich phenomenology, which makes it extremely predictive and testable.

A few previous accounts on the observational status of the Galileon model \cite{2013PhRvD..87j3511B, Barreira:2014jha} pointed out that ISW-related data plays a particularly powerful role in tests of this theory of gravity. An example is the low-$\ell$ part of the CMB temperature power spectrum, which places tight bounds on the values of the Galileon parameters, $c_i$, and constraints the expansion rate at late-times to follow the so-called tracker evolution (cf.~Eq.~(\ref{eq:hubletracker})). The CMB temperature power spectrum is, however, insensitive to the sign of the ISW effect which is positive if the lensing potential decays at late times (as it does in $\lcdm$), and negative if it gets deeper. In \cite{Barreira:2014jha}, the authors have noted that the lensing potential in the Galileon model can have non-trivial evolutions and argued qualitatively that data sensitive to the sign of the ISW effect (such as the cross-correlation of CMB temperature with foreground galaxies) may help put even tighter constraints, potentially ruling out this entire theory of gravity.

In this paper, we set out precisely to quantify the degree of tension (if any) between the Galileon model and ISW data. For the latter we considered the data from \cite{Ferraro:2014msa} for the cross-correlation of CMB temperature maps with the distribution of galaxies in the WISE survey. To carry out our investigation we have first performed a MCMC constraint analysis on the Galileon model using CMB temperature and lensing data from Planck and BAO data (called CMB+BAO13 dataset here, c.f.~\autoref{sec:mcmc}). Then, we have re-sampled the resulting Markov chains to inspect the corresponding ISW predictions and to see how they compare to the measured WISE ISW signal. In order to compute the cross-correlation between CMB temperature and WISE galaxies, $C_{\ell}^{{\rm Tg}}$ (cf.~\autoref{sec:ISW}), one must first fit for the bias of the WISE galaxies. We have done this by using the cross-correlation of the CMB lensing convergence maps with the WISE galaxies, $C_{\ell}^{\kappa {\rm g}}$ (cf.~\autoref{sec:bias}).

Our analysis steps were applied separately to the cubic, quartic and quintic sectors of the Galileon model (cf.~\autoref{sec:gal-bg}). The results can be summarized as follows:

\begin{enumerate}

\item Our constraints recover the fact that in Galileon cosmologies there is a strong preference for non-zero neutrino masses and that the resulting best-fitting values of $H_0$ are compatible with local measurements. Both these two aspects are very different than what happens in standard $\lcdm$ (cf.~\autoref{fig:xic3-H0}).

\item In the cubic Galileon model the amplitude of the WISE ISW signal is always negative within the regions of parameter space preferred by the CMB+BAO13 data. The degree of the tension is at the $7.8\sigma$ level which effectively rules out the simpler ``corner'' of Galileon gravity.

\item In Quartic and Quintic Galileons the WISE ISW signal also rules out a significant portion of the parameter space, but not all of it. For these more general Galileon models there are regions of parameter space that yield good fits to the CMB, BAO13 and ISW data (cf.~\autoref{fig:ISW_Gnu4}, \autoref{fig:ISW_Gnu5}).

\item The quartic and quintic Galileon models that ``survive'' the WISE ISW tests exhibit some tensions with recent BAO data. The significance of these tensions can reach the $1.5\sigma-2.4\sigma$ levels (cf.~\autoref{fig:baodist}, \autoref{tab:BAO_chi2}), which suggest that future BAO data may prove particularly powerful at constraining further Galileon gravity.
\end{enumerate}

\bigskip

Of the four bullet points above, the last three are new compared to the previous constrain analysis of \cite{Barreira:2014jha}.

A general prediction of the quartic and quintic models is that the sign of the ISW amplitude is a redshift-dependent quantity. While we found no tension of the Galileon models with ISW measurements from the NVSS galaxy sample that spans a broad redshift range (up to $z \sim 5$), data from galaxy samples in narrow redshift bands around between $z \sim 0-1$ could provide useful information to test these models further. A difficulty here, that is general to all ISW-related observables, is that they are only important on very large scales where the signal-to-noise of the data is significantly limited by cosmic variance.

Furthermore, future BAO measurements could increase the tension of the Covariant Galileon to the data to a confidence level that also rules out the quartic and quintic sector of the model. In typical BAO analysis the reconstruction of the density field of the galaxies plays a crucial role to reduce the error bars of the BAO measurements (see e.g.~\cite{reconstruction}). However, current reconstruction implementations have assumed GR. As statistics improve in BAO scale determinations it may be worthwhile to revisit the impact of assuming GR in the reconstruction procedure \cite{reconstruction} and what systematic biases (if any) this might introduce in constrains of modified gravity models%
\footnote{Non-linear effects on the BAO scale have been studied in the context of Galileon gravity, showing that both the BAO shift and the perturbation theory kernel can depart significantly from the standard prediction \cite{Bellini:2015oua}.}.

Compared to $\lcdm$, the Galileon model has many more distinct signatures that can be further probed with cosmological data. At the background level, in addition to future BAO data, future higher-precision model-independent determinations of $H_0$ will also help to distinguish between these competing cosmological models. At the level of large scale structure formation, the higher amplitude of the CMB lensing potential at low-$\ell$ (cf.~\autoref{fig:CMB-TT-pp}) can help place tight constraints in the Galileon model, potentially ruling it out if future and more robust analyses on the largest angular scales of the sky confirm the current trend of the data. The lensing signal associated with cosmic voids is also a potentially powerful way to further test these models. In \cite{2015JCAP...08..028B}, the authors demonstrated that for the Cubic Galileon the differences to $\lcdm$ are appreciable because screening effects are not at play in these under-dense regions. It would therefore be interesting to extend the analysis of \cite{2015JCAP...08..028B} to the more general quartic and quintic sectors (similar lines of reasoning apply to the lensing signal of galaxy troughs \cite{Gruen:2015jhr, 2017JCAP...02..031B}).

In future work it would also be interesting to check whether in Galileon cosmologies the CMB constraints on the $\Omega_m - \sigma_8$ plane are consistent with those coming from lensing shear data (see e.g.~\cite{2016arXiv161004606J} for an investigation of the impact of departures from $\lcdm$ in alleviating the tension that currently exists between these two datasets). The next generation of galaxy surveys can also prove useful in pinpointing the competing time- and scale-dependent effects of neutrino masses and the enhanced gravitational strength in the Galileon model (see e.g.~\cite{Alonso:2016suf,Bellomo:2016xhl}). Further, investigations of non-linear structure formation in these models (using e.g. the N-body code developed in \cite{Li:2013tda} for the Quartic Galileon) are also welcomed to better understand the phenomenology of Galileons and design novel observational tests in regimes where screening becomes important. Such efforts with N-body simulations should include modelling of massive neutrinos.

The covariant Galileon can also be tested by several non-cosmological observables, including gravitational waves, astrophysical tests, terrestrial neutrino experiments and local gravity experiments. In Quartic and Quintic Galileons the scalar field induces an anomalous speed in the gravitational wave propagation \cite{Lombriser:2015sxa,Brax:2015dma,Bettoni:2016mij}: the observation of gravitational wave events with electromagnetic counterparts would lead to a phenomenally precise test of the model using Earth or space-based detectors \cite{Bettoni:2016mij}. Galileon gravity can also be put to test by confronting the appreciable neutrino mass fractions preferred by cosmological data with future laboratory experiments that aim to be sensitive to sub-eV absolute mass values: the KATRIN experiment \cite{Osipowicz:2001sq,Drexlin:2013lha} will probe $\sum m_\nu \gtrsim 0.6$eV, while more into the future, Project 8 might reach sensitivities of $m_{\nu_e}\gtrsim 40$meV ($\sum m_\nu \gtrsim 0.1$eV) \cite{Esfahani:2017dmu}. Finally, as any modified gravity model, astrophysical, Solar System and laboratory tests \cite{Brax:2011sv} should all serve to test the validity of the model in complementary ways. Here, the challenges lie in designing experiments that are sensitive to the small values of the modifications to gravity in strongly screened regimes; or in a more robust understanding of the local value of $\dot{\phi}$ (cf.~footnote 9), which if sizeable could well rule out the quartic and quintic Galileon models by means of lunar laser ranging experiments \cite{Babichev:2011iz} or gravitational waves from binary pulsars \cite{Jimenez:2015bwa}.

All these considerations point out that Galileon gravity offers a testable and concrete {\it working case model} to help explore typical observational signatures of theories beyond GR in cosmological and non-cosmological set-ups.

\paragraph{Acknowledgements:}
 
We are very grateful to Simone Ferraro for helpful discussions and providing the ISW data used in this work. Furthermore we thank Jos\'e Luis Bernal, Ben Elder, Franz Elsner, Simone Ferraro, Elena Giusarma, Baojiu Li, Edvard M\"ortsell, Hiranya Peiris, Valeria Pettorino, Shun Saito, Iggy Sawicki, Fabian Schmidt, Anowar Shajib, Sunny Vagnozzi and Licia Verde for useful suggestions and comments on the draft.
J.R. acknowledges support by Katherine Freese through a grant from the Swedish Research Council (Contract No. 638-2013-8993). MZ is supported by the Maria Sklodowska-Curie Global Fellowship Project``NLO-CO''.

\bibliographystyle{JHEP}

\bibliography{refs}

\end{document}